# Using Machine Learning for Model Physics: An Overview

Vladimir Krasnopolsky[1*] and Alexei Belochitski[2]

[1]Environmental Modeling Center/National Centers for Environmental Predictions/National Weather Service

[2]I. M. Systems Group at Environmental Modeling Center/National Centers for Environmental Predictions/National Weather Service

[*)] Corresponding author email: Vladimir.Krasnopolsky@noaa.gov

# Abstract

In this overview, a generic mathematical object (mapping) is introduced, and its relation to model physics parameterization is explained. Machine learning (ML) tools that can be used to emulate and/or approximate mappings are introduced. Applications of ML to emulate existing parameterizations, to develop new parameterizations, to ensure physical constraints, and control the accuracy of developed applications are described. Some ML approaches that allow developers to go beyond the standard parameterization paradigm are discussed.

# 1. Introduction

The scientific and practical significance of complex interdisciplinary numerical models and prediction systems has increased tremendously during the last few decades, due to improvements in their fidelity via better understanding of the basic processes and their interrelationship (see Section II of this book), as well as development of new conceptual approaches and computational techniques (see Section III of this book), and in Earth observing system (see Section IV of this book). A well pronounced trend has been revealed in weather and climate numerical modeling. This trend marks a transition from investigating simpler linear or weakly nonlinear single-disciplinary systems like idealized atmospheric or oceanic models that include a simplified description of the physical processes, to studying complex nonlinear weather and climate systems such as coupled atmospheric-oceanic-land-ice systems that incorporate sophisticated representations atmospheric physics, chemistry, land-surface interactions, etc. Some of these models are initialized by complex data assimilation systems that utilize a very large number of observations from different sources and integrate these data with the forecasts generated by a large ensemble of lower resolution numerical models.

Due to their growing complexity (e.g., increase of horizontal and vertical resolutions, and/or the number of ensemble members, and increase in sophistication of model physics), both global and regional modeling activities consume a tremendous amount of computational resources, presenting model developers and users with a significant challenge despite expanding computing capabilities. Ensemble forecasting systems are particularly impacted by the limits on computational resources that constrain their resolution and/or the number of the ensemble members. Various approaches are being considered to alleviate this problem. For example, in the European Centre for Medium-Range Weather Forecasts (ECMWF) the modelers consider transitioning from double to single precision calculation to improve the performance of Integrated Forecasting System and to allow further increase in vertical resolution (Lang et al. 2021).

Despite significant progress in understanding of physical phenomena and improvement of observing systems, substantial uncertainties remain in representation of many processes in numerical models, e.g., effects of clouds in general circulation models of the atmosphere (e.g., Ceppi et al, 2017). New flexible and powerful numerical techniques are required to learn underlying physics from data that are available.

During the last decade, machine learning (ML) began to play an important role in advancing scientific discovery in domains traditionally dominated by physically based (first principle) models: in climate, weather, and earth system sciences (Krasnopolsky and Fox-Rabinovitz 2006a, Hsieh 2009, Krasnopolsky 2013, Reichstein et al. 2019, Wang et al. 2021), in oceanography in prediction of ocean weather and climate, habitat modelling and distribution, species identification, coastal water monitoring, marine resources management, detection of oil spill and pollution and wave modelling (Ahmad 2019), in solid Earth geoscience (Bergenet al. 2019), in astronomy (Izi´c et al. 2019), in fluid dynamics (Kutz 2017), in image processing (Wang et al. 2018, Ebert-Uphoff and Hilburn 2020), in biomedical sciences (Alber et al. 2019), in atomic physics (Brockherde et al. 2017), in high energy and fusion physics (Kasim et al. 2020), in engineering (Alizadeh 2020), etc. Such a broad application of ML is possible because problems encountered in these different fields are similar from the mathematical point of view. Many of them belong to various aspects of mapping problems (see next Section for description of mapping). The use of ML models is particularly promising in scientific problems involving processes that are not completely understood, or where it is computationally infeasible to run physically based models at desired spatiotemporal resolutions. Some of the most complex problems of this nature are posed by Numerical Weather/Climate Prediction Systems (NWPS).

Attempts to replace physically based models with the state-of-the-art "black box" ML-based ones have often met limited acceptance in scientific domains due to inability to provide a meaningful physical interpretation of underlying processes, or their large data requirements, and their limited generalizability to out-of-sample scenarios. Given that neither an ML-only nor a physically based-only approach can be considered sufficient for complex scientific and engineering applications, the research community has been exploring the continuum of hybrids of physically-based and ML-based models, where both scientific knowledge and data are integrated in a synergistic manner (Krasnopolsky and Fox-Rabinovitz 2006a, Rai and Sahu 2020). This paradigm is fundamentally different from mainstream practices in the ML community where domain-specific knowledge is often used in secondary roles, e.g., feature engineering or post-processing. In contrast to these practices that can only work with simpler forms of heuristics and constraints, this overview is focused on hybrid NWPSs incorporating a tighter coupling of ML-derived model components with scientific knowledge embedded into physically based models.

ML is being increasingly applied to reduce demands for computational resources (e.g., Alizadeh et al., 2020, Kasim et al., 2020) and to accelerate calculations in NWPSs in particular (Chevallier et al., 2000; Krasnopolsky et al., 2002; Krasnopolsky et al., 2005; Krasnopolsky, 2013, Gentine et al., 2018, Wang et al., 2019, Lagerquist et al., 2021). Another venue for application of ML in NWPSs is development of new ML-based parameterizations derived from observed and/or simulated data (Krasnopolsky et al., 2013; Schneider et al., 2017; Brenowitz & Bretherton, 2018; O'Gorman & Dwyer, 2018, Beucler et al, 2020; Watt-Meyer et al., 2021).

In Section 2 of this overview, a generic mathematical object (mapping) is introduced, and its relation to model physics parameterization is established. In Section 3, ML tools that can be used to emulate and/or approximate mappings are introduced. Section 4 describes the use of ML to emulate existing parameterizations. Section 5 discusses using ML to develop new parameterizations, Section 6 introduces the concept of ML ensembles, Section 7 describes methods to ensure physical constraints, Section 8 introduces ML tools allowing to go beyond the standard parameterization paradigm, and Section 9 summarizes the discussion.

## 2 Parameterizations as mappings

A mapping, $M$, is a relationship between two vectors $X$ (input vector) and $Y$ (output vector) that can be symbolically written as,

$$Y = M(X);\ X \in R^N\ and\ Y \in R^M \tag{1}$$

where $N$ and $M$ are the dimensionalities of the input and output spaces correspondingly. Any parameterization of model physics, and even the entire suite of physically-based parameterizations, as well as the model itself, is a mapping between a vector of input parameters (e.g., profiles of atmospheric state variables) and a vector of output parameters (e.g., a profile of heating rates in radiation parameterization). In terms of $Y$ vs. $X$ dependencies, parameterization mappings may be continuous or almost continuous (in other words, containing only a finite number of step function-like discontinuities). Usually, physically-based parameterizations do not contain singularities.

Another very important property of the mapping (1), relevant for parameterization mappings, is mapping complexity (Krasnopolsky, 2013; Chapter 2.2.2). There are many types of mapping complexity that are discussed in the literature (e.g., Alizadeh et al. 2020). When it comes to mappings defined by parameterizations, at least four different measures of mapping complexity can be suggested. The parameterization mapping, $M$ (1), is usually a symbolic representation of a mathematical formalism based on first principles and describing a chain of physical processes or hierarchy of interacting physical processes (e.g., atmospheric radiation). Therefore, we can talk about the *physical complexity* of the mapping (1) that corresponds to the complexity of the hierarchy of physical processes described by the mapping. We can also introduce quantitative or semi-quantitative characteristics of physical complexity: the number of physical processes involved and the number of levels (depth of the hierarchy) of the processes involved.

Also, we can consider *mathematical complexity* of the mapping (1) and introduce its quantitative or semi-quantitative measures: the number of equations, the type of equations (e.g., linear vs. nonlinear, ODE vs. PDE vs. integro-differential equations), and the dimensionality of the equations. It is noteworthy that the measures of physical and mathematical complexity are not unambiguous. For a particular physical process, alternative mathematical formalisms based on first principles often exist, leading to different types and numbers of equations describing the same physical system. As a result, several different estimates of physical and/or mathematical complexity may be obtained for the same mapping (1). Euler vs. Lagrange formulations of the equations of geophysical fluid dynamics and the Schrödinger vs. Heisenberg formulations of quantum mechanics are examples.

The third measure of complexity that can be introduced is *numerical/computational complexity* of a mapping, e. g. the number of elementary numerical operations required for calculating $Y$ given $X$. Numerical/computational complexity of a parameterization is closely related to the amount of computational resources required for its application. However, this measure is also ambiguous because different numerical schemes applied to the same set of equations (e. g., finite differences vs. variational methods for solving PDEs) may lead to dramatically different counts of elementary numerical operations. Here again, several different estimates of numerical complexity may be obtained for the same mapping (1).

The fourth type of mapping complexity is called *functional complexity*. It describes the complexity of the functional dependency of the outputs, $Y$, on the inputs, $X$, or the "smoothness" of this dependency. If the three previous definitions in some respects depend on, or are conditioned by, our knowledge of the internal structure of the target mapping (1), this fourth definition characterizes the complexity of the mapping as a whole, as a single elementary object that transforms the input vector into the output vector. It is intuitively clear that the functional complexity of the mapping (1) can, in principle, be measured unambiguously. Approximation of a parameterization mapping with ML tools, for example neural networks (NN), can provide a quantitative measure of the mapping functional complexity. For example, complexity of NN emulation (see eq. 7 and 8 below) can be used as a measure of the functional complexity of the parameterization mapping.

The functional complexity of a mapping is the most important measure when application of ML tools to mapping approximation is involved. For example, a parameterization with very high physical and/or computational complexity may have low functional complexity, e. g. a scheme composed of many complex equations may result in a comparatively simple output vs. input ($Y$ vs. $X$) dependence. A significant increase in computational performance achieved when physically-based parameterizations are replaced with their ML-based emulators suggests that the functional complexity of the real world physics schemes is lower than their physical and/or computational complexities. This situation is optimal when the goal of the ML application is to accelerate calculations because the computational burden in terms of required computer resources of an ML-based emulator is proportional to the functional, not physical or computational complexity, of the parameterization being emulated.

In some cases (e.g., stochastic physics or superparameterization) the parameterization mapping contains an internal source of stochasticity. It may be due to several reasons: a stochastic process that the mapping describes, a stochastic method (e.g., Monte Carlo methods) implemented in mathematical formulation of the mapping, contribution from the subgrid processes, or uncertainties in the data that are used to define the mapping. In this case we can modify the symbolic representation of the mapping (1) as,

$$Y = M(X, \varepsilon);\ X \in R^N\ and\ Y \in R^M \qquad (2)$$

where $\varepsilon$ is a stochastic vector variable explicitly representing the stochastic nature of the mapping, or the mapping uncertainty. A stochastic mapping given symbolically by (2) is a family of mappings sampled from a distribution $Y = M(X, \varepsilon)$. The uncertainty $\varepsilon$ is an inherent part of the stochastic mapping, which determines its range, domain and shape.

The parameterization mapping may be defined explicitly or implicitly. It can be defined explicitly as a set of equations based on first principles and/or empirical dependencies (e.g., radiative transfer or heat transfer equations) or as a computer code. A collection of data records (e.g., observations, measurements, computer simulations) represents the parameterization mapping implicitly.

In what follows, an ML technique will be used to approximate (or emulate) existing physically based parameterizations. We will call the resulting ML-based component or ML emulation of existing parameterization. Both of these terms will be used interchangeably. ML can be used to develop entirely new ML parameterizations (MLP) based on learning from the observed or/and simulated data. Both applications are possible due to the fact that the ML techniques (e.g., neural networks (NN), support vector machines (SVM), or decision trees) are mappings that can be used to approximate mappings given by parameterizations or collections of data records. A more detailed description of mappings and their properties can be found in Krasnopolsky (2013, Chapter 2.2).

## 3 Machine Learning tools to approximate parameterization mappings

ML is a subfield of artificial intelligence that uses statistical techniques to give computers the ability to "learn" (i.e., progressively improve performance on a specific task) from data, without being explicitly programmed (Bishop, 2006). This definition explains why ML is sometimes also called statistical learning or learning from data (Cherkassky & Muller, 1998). The set of ML tools includes a large variety of different algorithms such as NNs, different kinds of decision trees (e. g., random forest (RF) algorithm), kernel methods (e. g., SVM and principal component analysis), Bayesian algorithms, etc. (see Figure 1). Some of these algorithms are more universal (e. g., generic multilayer perceptron or NNs), some of them are more focused on a specific class of problems (e. g., convolutional NNs that show an impressive performance as image/pattern recognition algorithms).

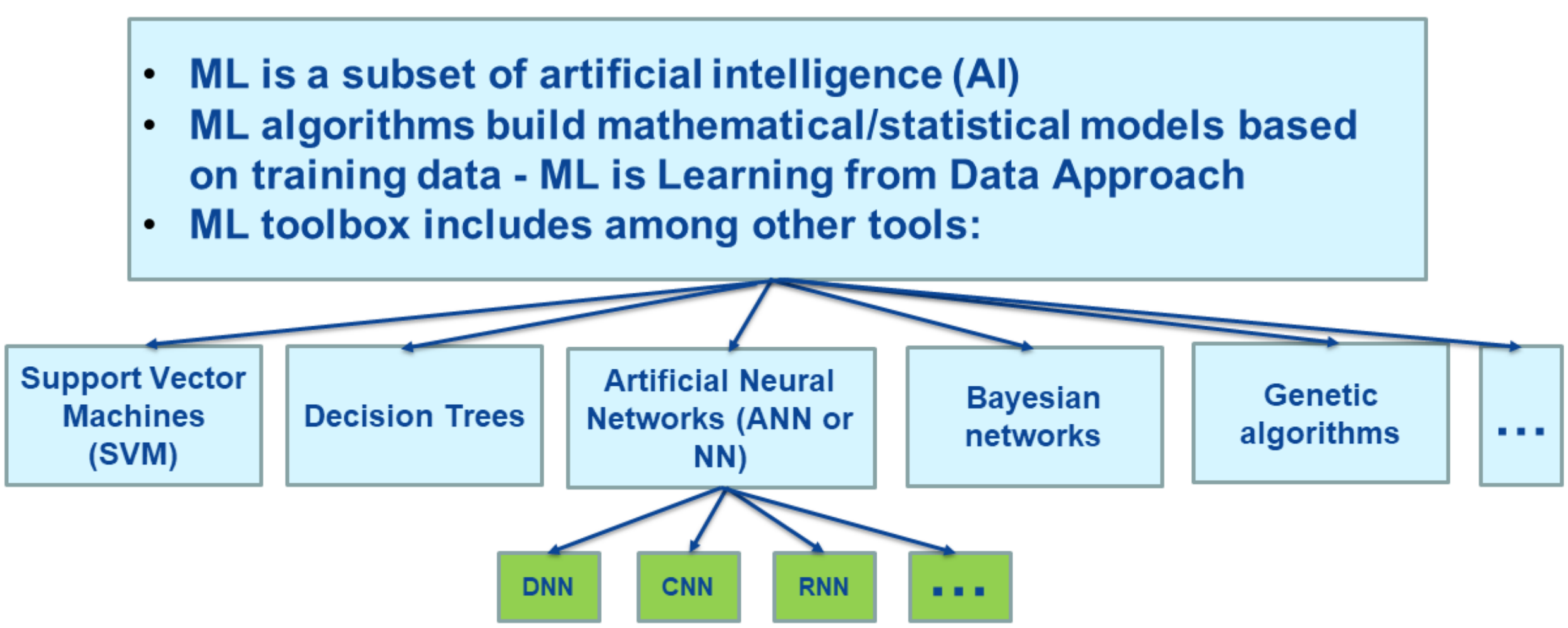

**Figure 1.** ML and various types of ML tools. New ML tools emerge very often.

To the best of our knowledge, two major types of ML tools that have been applied to model physics parameterization mappings are NNs (e.g., Chevallier et al., 2000; Krasnopolsky et al., 2002; Brenowitz & Bretherton, 2018; Rasp et al., 2018, Rasp, 2019, Beucler et al, 2020; Geer 2021; Xu et al, 2021) and tree algorithms (e.g., Belochitski et al., 2011; O'Gorman and Dwyer, 2018). There are many different types of NNs: shallow, deep, convolutional, recurrent, etc., as well as many types of tree algorithms. Belochitski et al (2011) showed that, when using a shallow (see the definition below) NN (SNN) to emulate an existing radiation parameterization, the NN performs better than tree algorithms in terms of accuracy and speedup. O'Gorman and Dwyer (2018) have found that when attempting to derive a new parameterization of subgrid processes based on a SNN, NN did not conserve energy and was slightly less accurate than a tree algorithm. Also, NN was not stable in prognostic single-column integrations. In their work, they experimented with using the simplest shallow, single hidden layer, NNs with a limited number of neurons (60 neurons). Krasnopolsky et al (2008a, 2010) and Belochitski and Krasnopolsky (2021) have shown that, when the accuracy of approximation is sufficient, the SNN emulation preserves invariants (e.g., energy or moist enthalpy) with an accuracy, sufficient for long term integration of the NN in the model without accumulation of errors and resulting instability. The stability in the process of integrating NN in the model could also be achieved by using a loss function that involves multiple time steps (Brenowitz & Bretherton, 2018). Using multiple time steps helps NN learn temporal continuity during the training, which is important when the NN is used in the online mode in the model. Introducing physical constraints into the loss function also results in stable and accurate integrations of the hybrid model (Beucler et al, 2020). Vast majority of published work on applying ML to model parameterization mappings used various types of NNs. The few works that used tree algorithms did not show significant advantages as compared to NNs. In the following we focus on properties of NNs and NN applications to model parameterization mappings. A limited discussion of other ML algorithms is given in the Summary.

### 3.1 Neural Networks

A generic NN that is used for modeling/approximating nonlinear multidimensional mappings is called the multilayer perceptron. It is comprised of "neurons" that are arranged in "layers". A generic neuron can be expressed as,

$$x_j^{n+1} = \phi\left(b_j^n + \textstyle\sum_{i=1}^{k_n} a_{ji}^n \cdot x_i^n\right) \tag{3}$$

Eq. (3) represents a neuron number *j* in the layer number *n*. $x_j^{n+1}$ is the output of the neuron that, at the same time, is an input to neurons of the layer number *n+1*. Here $x_j^n$ are inputs to neurons of the layer number *n* (outputs of neurons of the layer number *n-1,* the input layer corresponds to *n=0*), *a* and *b* are fitting parameters or NN weights and biases, $\varphi$ is the so-called activation function, and $k_n$ is the number of neurons in the layer number *n*. The entire layer number *n* can be represented by a matrix equation:

$$X^{n+1} = \phi(B^n + A^n \cdot X^n) \tag{4}$$

where for $n = 0$, $X^n = X$, a vector of the NN inputs If the layer number *n+1* is the output layer, linear neurons are often used,

$$y_j = x_j^{n+1} = b_l^n + \sum_{i=1}^{k_n} a_{ji}^n \cdot x_i^n$$

or for the entire output layer,

$$Y = X^{n+1} = B^n + A^n \cdot X^n \qquad (5)$$

here $Y$ is a vector of outputs.

The activation function $\varphi$ is a nonlinear function, often specified as the hyperbolic tangent; however, rectangular linear unit, softmax, leaky rectangular linear unit, Gaussian, trigonometric functions, etc. are also used (e.g., Hsieh 2009).

All layers of multilayer perceptron NN between input and output layers are called “hidden layers”. NNs with multiple hidden layers are called “deep neural networks” (DNN). The simplest multilayer perceptron NN with one hidden layer is called a “shallow” NN (SNN),

$$Y = B^1 + A^1 \cdot \phi(B^0 + A^0 \cdot X)$$

Or

$$y_j = b_j^1 + \sum_{i=1}^{k} a_{ji}^1 \cdot t_i \ , \ \ j = 1, \cdots, M$$

$$t_i = \phi\left(b_i^0 + \sum_{s=1}^{N} a_{is}^0 \cdot x_s\right) \qquad (6)$$

The SNN is a generic analytical nonlinear approximation or model for the parameterization mapping (1) and a mathematical solution of the machine learning problem (Vapnik 2019).

The expansion (6) isa linear combination of nonlinear neurons, or basis functions, $t_j$, where the coefficients $a_{ji}^1$ ($j = 1,\ldots,M$ and $i = 1,\ldots,k$) are the linear coefficients of this expansion. It is essential that the basis functions $t_j$ (6) are nonlinear both with respect to inputs $x_s$ ($s = 1,\ldots,N$) and the fitting parameters, or coefficients, $a_{is}^0$ and $b_i^0$ ($i = 1,\ldots,k$). Due to the nonlinear dependence of the neurons on multiple fitting parameters $a_{is}^0$ and $b_i^0$, $\{t_i\}_{i=1,\ldots,k}$ is a flexible and versatile set of non-orthogonal basis functions that adjusts to the functional complexity of the approximated mapping (1): it has been shown by multiple authors in a variety of contexts that the family of functions (6) can approximate any continuous or almost continuous (with a finite number of finite discontinuities) mapping (1) (Cybenko, 1989; Funahashi, 1989; Hornik, 1991; Chen & Chen, 1995). Vapnik (2019) proved that for a wide set of functions in the Hilbert space, the solution that minimizes Empirical Risk functional in a set of functions with a bounded norm defines a "shallow network" a neural network with just one hidden layer). The accuracy of the SNN approximation or the ability of the SNN to resolve details of the parameterization mapping (1) is proportional to the number of basis functions, *k* (Attali & Pagès, 1997).

The NN (3,4) itself belongs to the class of mappings given by (1). Both the *computational and functional complexity* of an NN can be defined as the number of fitting parameters *a* and *b* in (3,4) (Krasnopolsky, 2013) . Therefore, for a deep neural network the complexity, is given by

$$\mathbb{C} = \sum_{i=0}^{n} k_{i+1}(k_i + 1) \tag{7}$$

where $n$ is the number of hidden layers in the DNN, $k_i$ is the number of neurons in the hidden layer number $i$, $k_0 = N$ is the number of inputs, and $k_n = M$ is the number of outputs.

For an SNN the equation 7 can be simplified:

$$C = k \cdot (N + M + 1) + M \tag{8}$$

Different SNNs approximating the same target mapping (1) with a fixed number of inputs *N* and outputs *M*, differ in complexity proportionally to the number of basis functions, *k*, used in individual NNs. The NN complexity grows linearly with the growth of the dimensionalities of both the input space (the number of inputs), *N,* and the output space (the number of outputs), *M*. The number of NN weights per output,

$$c = \mathbb{C} / M \tag{9}$$

reflects the computational and functional complexity of the dependency of each NN output vs. NN inputs. This measure of the NN complexity is useful when NNs with different numbers of outputs are compared.

A pictographic language reminiscent of a data flow chart is used traditionally in the NN field. In this language, a basis function $t_j$ (4), or neuron is represented as shown in Figure 2 (a and b). The neurons are arranged into *layers* inside the NN (Fig 2c shows an SNN). The input layer is different from other layers of the NN: input neurons do not perform any numerical operations; instead, they simply provide input data to neurons in the following hidden layer.

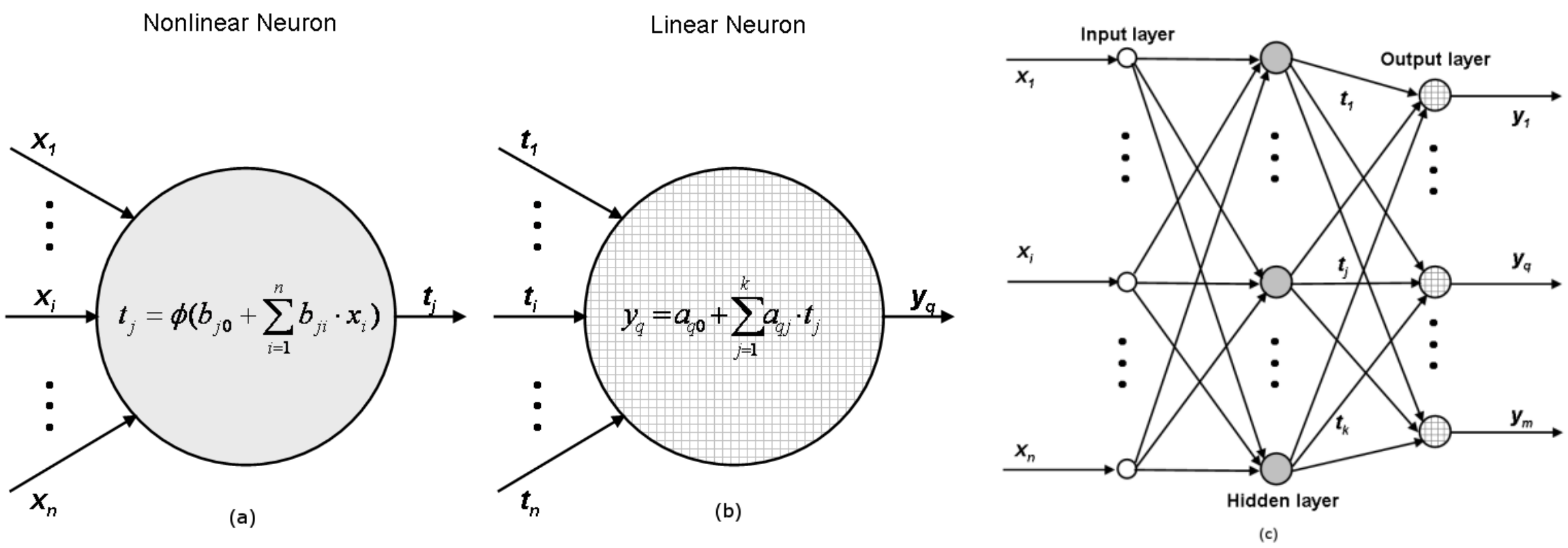

**Figure 2.** The figures show: (a) nonlinear neuron, (b) linear neuron (eq. 13.4), and (c) - the simplest SNN with one hidden layer and linear neurons in the output layer (eq. 13.6).

Hidden layers are generally composed of nonlinear neurons (eq. 4 and Fig. 2a). The neurons in the output layer are usually linear (Fig. 2b). The connections (arrows) in Figure 2 correspond to the NN weights, the name used for the fitting parameters *a* and *b* in NN jargon.

For the simplest SNN considered shown in Fig. 2c, there is a one-to-one correspondence between eqs. (6) and Fig.2c. However, in general (e.g., for more complex types of NNs), the pictographic language (Fig.2) is not redundant. This language can suggest NN architectures that probably cannot be represented analytically in terms of a closed set equations (e.g., some convolutional and recurrent NNs, NNs with missed connections, complex NNs that are composed of several NNs of different types). The pictograms that represent the design of such NNs cannot be described by a closed set of equations; however, these pictograms can be translated into computer codes.

As mentioned above, the SNN architecture is sufficient for the approximation of any continuous (or almost continuous) mapping. Additional hidden layers (DNN) and/or nonlinear neurons in the output layer can be introduced and the resulting NN applied to either mapping approximation problem or problems of different nature, for example, nonlinear principal component analysis (Hsieh, 2001), extreme event prediction (e.g. forecasting large hail) (Gagne et al., 2019), and pattern recognition (e.g. detection of synoptic-scale fronts) (Lagerquist et al., 2019). DNNs have been extremely successful in many areas like pattern recognition, classifications, image processing, etc. However, as pointed out by Vapnik (2019), from the standpoint of statistical learning theory, only the SNN has been formally shown to be a solution to the mapping approximation problem. Successful approximation of the mapping (1) by a DNN cannot be guaranteed, and this specific application of DNNs should be considered a heuristic approach. Both SNNs and DNNs have been applied to parameterization mappings by different authors (also see discussion in Section 9.3).

### 3.2 Training set

In a practical application, a parameterization mapping (1) is usually represented and presented to the NN by a data or training set, $C_T$, that consists of $P$ pairs of input and output vectors $X$ and $Y$,

$$C_T = \{X_p, Y_p\}_{p=1,\dots,P} \tag{10}$$

where $Y_p = M(X_p) + \xi_p$, $X_p \in D$ and $Y_p \in R$, $\xi_p$ represents any errors associated with the observations or calculation with a probability density function $\rho(\xi)$, $D$ and $R$ are the mapping domain and range respectively. The set $C_T$ can also be considered as a combination of two rectangular matrices,

$$C_T = \{C_X, C_Y\}, \tag{11}$$

where $C_X$ is a matrix of dimensionality $N \times P$ composed of all input vectors $X$, and $C_Y$ is a matrix of dimensionality $M \times P$ composed of all output vectors $Y$. The training set is all that the NN "knows" about the parameterization mapping that it is expected to approximate. This is the reason why it belongs to a class of *data driven* or *learning from data* methods (Cherkassky & Mulier, 2007).

The training set represents the mapping (1) for the NN and, therefore, it needs to be *representative*. It means that the training set must have a sufficient complexity to represent the complexity of the target mapping (1), allowing the NN to achieve the desired accuracy of the approximation of the target mapping. The set should have a sufficient size, $P$, of properly distributed data points that adequately resolve the functional complexity of the parameterization mapping (1). The training set should have finer resolution where the target mapping is not smooth and coarser resolution where it is smoother, namely, the domain $D$ should be properly sampled. If the domain is undersampled, the training set is not representative. There are no generic recipes of how to determine if the domain is undersampled. In each particular case this problem should be solved taking into account spatial and temporal scales of underlying physical processes, the model resolution, and the frequency and spatial and temporal distributions of extreme events.

Understanding the configuration of the mapping domain and its properties is essential for any application of the mapping (1) and for proper NN training and application (Bishop, 1995). If all components of the input vector $X$ are scaled to the range [-1.,1.], the volume of the input space $R^N$ is equal to $2^N$ and, therefore, grows exponentially with $N$. In a numerical model, once the space is discretized on a grid by $K$ values per dimension, then the problem grows even faster, as $K^N$. It means that in the input space we have $K^N$ grid cells, and to represent our mapping, we need an exponentially large training set in order to ensure that each grid cell contains at least one data point. This exponential growth in the amount of data with the increase of the input space dimensionality is often called the *curse of dimensionality* (Bishop, 1995; Vapnik & Kotz, 2006).

Fortunately, for parameterization mappings, the input vector *X* consists of atmospheric state variables that are related and correlated due to multiple physical and chemical processes in the atmosphere. The inter-relations and correlations between inputs reduce the size and the effective dimensionality of the domain. Thus, the mapping domain, *D*, spanned by the input vectors *X* occupies only a part of the input space $R^N$. As a result, for parameterization mappings, the mapping domain, *D*, may be significantly smaller than an *N*-dimensional cube in the input space $R^N$ . This may significantly simplify (in many cases make possible) the sampling task, especially for cases of high input dimensionality, helping to avoid the curse of dimensionality. In general, it is often very difficult, if possible at all, to determine the actual shape, location, and effective dimensionality of *D*, which makes it difficult to adequately sample the mapping domain *D*. However, for parameterization mappings, when observed data or data simulated by an atmospheric model are used, the nature or model automatically, by default, place the data inside the domain *D*. Also, understanding relationships between components of the input vectors *X* allows to reduce dimensionality of the input vector, removing redundant components (Krasnopolsky et al., 2009; Krasnopolsky et al., 2010). For example, for the short-wave radiation parameterization in NCEP Climate Forecast System (CFS), the input vector *X* has more than three thousand components. By removing the redundant components, the dimensionality of *X* was reduced to $N = 562$. Even after such a tremendous reduction of dimensionality, generally speaking, $P > 2^{562}$ data points are necessary to sample the input space and create a training set. However, due to the use of data simulated by CFS, which samples only the parameterization domain *D*, the size of the training set could be reduced to a reasonable size, $P = 2 \times 10^5$ (Krasnopolsky et al., 2010). When MLPs are developed, simulated data or observations fused with the simulated data (analysis) are used. Usually there is no shortage of the training data. The only problem here may be the big size of the training set (especially if DNNs are used) that may increase training time and memory requirements even beyond capacities of modern high-performance computing (HPC) systems.

Another important consideration is a relationship between the size of the training set, the complexity of the NN architecture, and concepts of underfitting and overfitting. Fig. 3 demonstrates some important issues in this regard. The orange squares represent the data set that was generated using a parabola ($y = a \cdot (x + b)^2$) with some noise added. Fig. 3a shows an approximation of the data using a minimal complexity linear model. The linear approximation completely misses the major feature of the data - the quadratic dependence of $y$ on $x$. Thus, the data are underfitted; the model is too simple; the complexity of the approximation model is lower than the complexity of the data set.

Fig, 3c shows the opposite situation. The complexity of the selected approximation model is significantly higher than the complexity of data; the model overfits the data (fits the noise). Fig. 3b shows the case when the complexity of the model corresponds to the complexity of the data set.

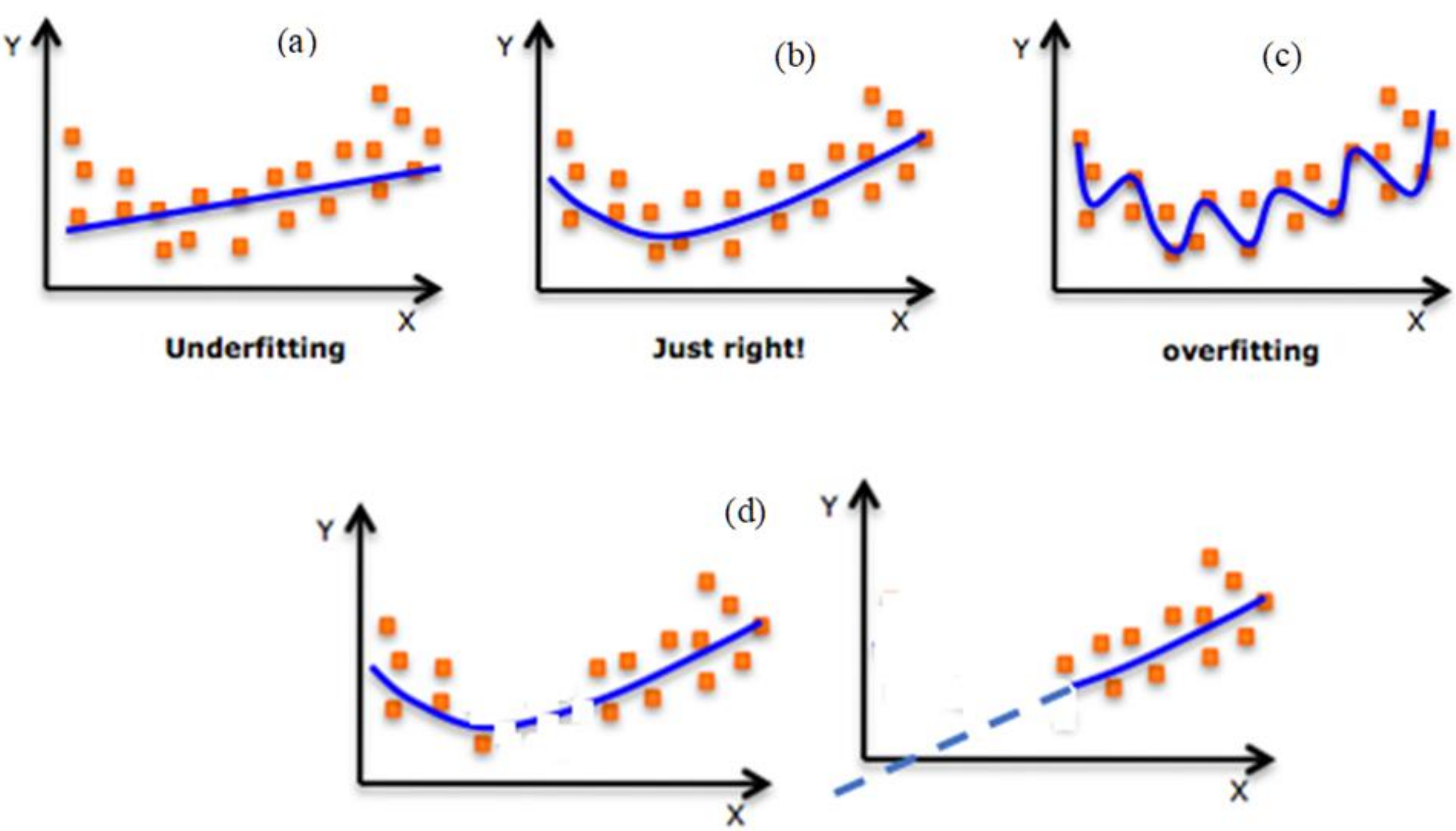

**Figure 3.** Relationship between complexity of approximating function and underfitting and overfitting (see in the text).

Figs.3d illustrates cases when the data set is incomplete, and a generalization is required. In the first case (left panel) successful generalization (interpolation) is performed. In the second case generalization fails because the data set is not representative (the entire left corner is missed) and the model is forced to extrapolate; however, the nonlinear extrapolation is an ill-posed problem and, in this case, extrapolation is erroneous.

### 3.3 ML modelling parameterizations, special requirements

ML can be applied to parametrization at least in two different ways (Krasnopolsky, 2013): (1) as an *emulation technique* for *accelerating calculation* of parameterizations previously developed based on approximate description of underlying physical processes (e.g., NN emulations of radiation parameterizations by Krasnopolsky et al., 2010) and (2) for *developing new* "empirical" *parameterizations* based on observed data or data simulated by high resolution models in situations when underlying physical processes are very complicated and not fully understood (e.g., cloud physics, Krasnopolsky et al., 2013; Brenowitz & Bretherton, 2018; Beucler et al, 2020). The great flexibility of ML tools also allows for the combination of these two approaches. Thus, the terms an "emulating ML" or an "ML emulation" of the parameterization mapping (1) and "ML parameterization" are used here to distinguish between these two problems.

It is noteworthy that both aforementioned tasks are different from a typical ML problem, where minimizing the error function for the process in consideration is the end goal. The ML tool is trained on a training set and validated on a test set. If approximation errors on both sets are sufficiently small, the problem is solved and a ML tool is ready to use. In the case of applying ML to model parameterization mappings, validation on an independent data set is a necessary but not sufficient step. The developed MLP is a subsystem/component of larger systems (atmospheric, oceanic, or coupled models). After introducing MLP in the larger system, the larger system becomes a hybrid model, which is a combination of physically based components (e. g., model dynamical core, other parameterizations) with the MLP. In the hybrid model, MLP can behave differently than on the test set, even when test set approximation errors are small, because of two reasons: (i) The larger systems (e. g. GCM) are nonlinear complex systems; in such systems, small systematic, and even random, approximation errors can accumulate over time and produce a significant impact on the quality of the model results, and (ii) erroneous behaviors and instability can arise if the system travels to the boundaries of the training set domain. At the same time, Pincus et al. (2003) have demonstrated that unbiased, random, uncorrelated errors in radiative heating rates with magnitudes as large as the net cooling rate do not statistically affect forecast skill in a state-of-the-art weather forecasting model. Pincus and Stevens (2013) explain this result in the following way: random small local heating rate errors in the bulk of the atmosphere lead to local small-scale instabilities that are mixed away by the flow. They caution, however, that there is no such mechanism for the surface variables, such as skin temperature, and errors in surface fluxes can be more consequential. Therefore, on the one hand, the development and application framework for MLPs should be focused on obtaining both the highest possible approximation accuracy and the highest possible accuracy of MLP in hybrid model simulations. On the other hand, in some applications the hybrid model can tolerate approximation errors in the MLP component as long as they remain random, uncorrelated, and unbiased.

Thus, when a MLP is developed, in addition to the criterion of small approximation errors at least three other criteria are used: (i) the NN complexity, $C$, (7, 8) is controlled and restricted to a minimal level sufficient for good approximation and interpolation; (ii) independent training, validation, and test data sets are used; the validation set is used in the process of training to control overfitting and the test set is used after the training to evaluate interpolation accuracy; redundant training set (additional redundant data points added "in-between" training data points sufficient for a good approximation) may be used for improving the NN interpolation abilities; and (iii) the most important validation criterion for MLP is obtaining both the highest possible approximation accuracy and the highest possible accuracy of MLP in GCM simulations.

## 4 ML emulations to speed up calculation of parameterizations

Here we discuss the development of NN emulations for the model physics components. Thus, the terms an "emulating ML" or a "ML emulation" of the parameterization mapping (1) are used here to emphasize the differences between developing a ML emulation of an existing physically based parameterization and developing a new MLP based on learning from data simulated by finer resolution models and/or observations. In the case of a complex high resolution multicomponent modern model, calculations of model physics usually take a significant part of the total model computation time, especially if physics or a particular parameterization are calculated at each grid point and time step. Several different approaches are currently used to reduce the cost of these calculations. For example, radiative transfer parameterizations supply their host model with broadband fluxes and heating rates, which are obtained by integration over time, space, and frequency. Therefore, a trade-off between accuracy and computational expense can be found in how finely these dimensions are discretized (Hogan et al, 2017).

- *Discretization in time.* All GCMs update their radiative heating/cooling rates less frequently than the rest of the model fields. For example, National Centers for Environmental Prediction (NCEP) Global Forecast System (GFS) v16 general circulation model (GCM) in its operational configuration updates its radiative fields once per model hour, while updates to temperature, moisture, most cloud properties etc. due to unresolved physics processes happen every 150 model seconds, or 24 times per single radiation call. Updates due to dynamical processes happen even more frequently, every 12.5 seconds (Kain et al, 2020). This approximation is good for slowly changing fields of certain radiatively active gases but is less justified for small-scale clouds with lifetimes of an hour or less.
- *Discretization in space.* Some GCMs calculate radiative fields on a coarser spatial grid and interpolate them onto a finer grid used for the rest of the model variables. For example, the radiation grid in the European Centre for Medium-Range Weather Forecasts (ECMWF) Integrated Forecast System (IFS) v43R3 in the ensemble mode is 6.25 times coarser than the physics grid (Hogan et al, 2017). This may cause 2m temperature errors in areas of surface heterogeneity, e. g. coasts (Hogan & Bozzo, 2015).
- *Discretization/sampling in frequency space.* The Rapid Radiative Transfer Model (RRTMG), a parameterization of radiative transfer for GCMs used in NCEP GFS and ECMWF IFS, utilizes 14 bands in the short wave (Mlawer et al, 1997), while the parameterization used at United Kingdom Met Office Unified Model utilizes 6 (Edwards & Slingo, 1996). Monte Carlo Spectral Integration (Pincus & Stevens, 2009) performs integration over only a part of the radiative spectrum, randomly chosen in each point in time and space, allowing to increase temporal/spatial resolution of radiation calculations. Monte Carlo Integration of the Independent Column Approximation (McICA) (Pincus et al, 2003), integrates over the entire spectrum, but samples subgrid-scale (SGS) cloud properties in a random, unbiased manner, in each grid column in time and space, instead of integrating over them.

These techniques reduce horizontal, or vertical, or the temporal variability of the radiation fields, reducing the accuracy of a model's radiation transfer calculations and spatial or/and temporal consistency of radiative fields with other parts of the model physics and dynamics, which may, in turn, impact the accuracy of climate simulations and weather predictions. More frequent calculations of model physics, and future introduction of more sophisticated parameterizations will result in further increase in computational burden imposed by model physics.

Another approach currently used to reduce the cost of model physics calculations is the use of simplifying physical and mathematical approximations when developing parameterizations. For example,

- In the wind wave model (Tolman, 2009), the calculation of nonlinear wave-wave interactions (Hasselmann & Hasselmann, 1985; Hasselmann et al., 1985) requires roughly $10^3$ to $10^4$ times more computational effort than all other aspects of the wave model calculations. Present operational constraints require that the computational effort to estimate nonlinear wave-wave interactions should be of the same order of magnitude as for the rest of the wave model. This requirement was met with the development of the Discrete Interaction Approximation (DIA, Hasselmann et al., 1985). Two decades of experience with the DIA in wave models has identified DIA's significant shortcomings (Tolman et al., 2005).
- The approach, originally labeled as a "super-parameterization" (SP), but later as the Multi-scale Modeling Framework (MMF), couples cloud-scale and large-scale dynamics within a single modeling framework (Khairoutdinov & Randall, 2001). The MMF replaces parameterizations of deep convection and moist processes with a 2D cloud resolving model (CRM). It is noteworthy that, even though 3D cloud dynamics is not considered, MMF still increases the run time of the host global model by a factor of 200 to 250, which severely limits its applicability (Wang et al., 2011)

- State-of-the-art modeling of microphysical cloud processes is tremendously computationally intensive and most atmospheric models include it in parameterized form (e.g., Morrison et al., 2020). Microphysical parameterizations introduce substantial simplifications and are sometimes derived for specific atmospheric scenarios while being applied globally. Even in parameterized form, microphysics calculations are computer resources and time consuming.

The aforementioned situation provides the motivation to look for alternative, faster, and accurate ways of calculating model physics, chemistry, hydrology and other processes. During the last two decades, an ML approach based on NN approximation or emulation was used for the accurate and fast calculation of atmospheric radiative processes (Krasnopolsky, 1996, 1997; Chevallier at al., 1998) and for emulation of model physics parameterizations in atmospheric numerical models (Krasnopolsky et al., 2002, 2005, 2008a, 2010). In these works, calculation of the model physics components has been accelerated by factors of 10 to $10^5$ as compared to the corresponding original parameterizations.

### 4.1 NNs emulating radiative parameterizations

Radiation parameterizations are one of the most computationally demanding parts of model physics. Approaches formulated by Chevallier at al. (1998, 2000) and Krasnopolsky et al. (1996, 2002, 2005, 2010) represent two different ways of hybridization of first principles and NN components in the physics parameterizations as well as in complex climate and NWP models. These approaches introduce hybridization at two different system levels: at the level of the subsystem (e.g., a single parameterization), and at the level of the entire system (e.g., numerical model). These two approaches lead to the concepts of a hybrid parameterization (HP) (Chevallier et al., 1998, 2000) and a hybrid model or hybrid GCM (HGCM) (Krasnopolsky et al., 2002, 2005; Krasnopolsky and Fox-Rabinovitz, 2006a,b; Goldstein and Coco, 2015).

Chevallier et al. (1998, 2000), following the physical structure of the long wave radiation (LWR) parameterization, a component of the ECMWF global atmospheric model, considered it as a combination of fluxes at vertical levels, $F_i$:

$$F(S,T,V,C) = \sum_i \alpha_i(C) \cdot F_i(S,T,V) \tag{12}$$

where $i$ is an index for the vertical level, the vector $S$ represents surface variables, $T$ is a vector (profile) of atmospheric temperatures, $C$ is a profile of cloud variables, and the vector $V$ includes all other variables (humidity profile, different gas mixing ratio profiles, etc.). Each partial or individual flux $F_i(S,T,V)$ is a continuous mapping and all discontinuities related to the cloudiness are included in $\alpha_i(C)$. In their hybrid parameterization, referred to as the "NeuroFlux", Chevallier et al. (1998, 2000) combined first principles based calculations of cloudiness functions $\alpha_i(C)$ with NN approximations for a partial or individual flux $F_i(S,T,V)$. As a result, the "NeuroFlux" hybrid LWR parameterization developed by Chevallier et al. (1998, 2000) is an array or battery of about 40 NNs. The number of required NNs (and speedup of calculations) depends on the number of vertical layers in the GCM, and at a vertical resolution of 60 layers or more, both sufficient accuracy and speedup of NeuroFlux cannot be achieved simultaneously (Morcrette et al., 2008).

Krasnopolsky et al. (2002, 2005) developed an emulation approach that considers the entire parameterization as a mapping (1), which is emulated using either a single NN or an ensemble of NNs. This ML emulation approach is based on the fact that any parameterization of model physics can be considered as a continuous or almost continuous mapping. In this approach, the desired accuracy of the NN emulation is achieved simultaneously with a significant speedup (see Table 1). Evaluation of accuracy of the NN emulation, as the first step, includes calculation of approximation errors by applying the NN emulation to an independent test data set. This step is necessary to identify NNs that are not sufficiently complex (approximation errors are significantly larger than errors on the training set). However, it is practically impossible to determine if the approximation errors are "sufficiently small" without the second crucially necessary step: interactive coupling of the NN emulation to its host model. Only evaluation of NN emulations in parallel model runs allows us to conclude that the approximation errors are sufficiently small, that these errors do not accumulate during the hybrid model run and have almost negligible impacts on the hybrid model behavior. It was demonstrated by Krasnopolsky et al. (2008a, 2010, 2012a) for NCAR CAM and NCEP CFS and GFS that approximation errors presented in Table 1 are sufficiently small.

In terms of the accuracy statistics presented in Table 1, there are practically no differences between NCAR CAM with 26 vertical layers and NCEP CFS with 64 vertical layers. Thus, the accuracy of the NN emulation approach does not depend significantly on the vertical resolution of the model. This fact illustrates the robustness of the NN emulation approach with respect to the changes in the model vertical resolution.

**Table 1.** Statistics for estimating the accuracy of the heating rate calculations (in K/day) and the computational performance (speedup) of the NN LWR emulation vs. the original parameterization for NCAR CAM (T42L26) and NCEP CFS (T126L64) LWR and SWR parameterizations. Total statistics show the bias, RMSE, the NN complexity $C$ (13.7) and average speedup[1] $\eta$. RRTMG is the Rapid Radiative Transfer Model for GCM.

| | **Statistics** | **NCAR CAM (L = 26)** | | **NCEP CFS (L = 64)** | |
|---|---|---|---|---|---|
| | | **LWR** | **SWR** | **RRTMG LWR** | **RRTMG SWR** |
| **Total Error Statistics** | **Bias** | $3.\cdot 10^{-4}$ | $6.\cdot 10^{-4}$ | $2.\cdot 10^{-3}$ | $5.\cdot 10^{-3}$ |
| | **RMSE** | 0.34 | 0.19 | 0.49 | 0.2 |
| **NN Complexity** | ***c*** | 490 | 810 | 520 | 706 |
| **Speedup, $\eta$** | ***Times*** | 150 | 20 | 16 (20) | 60 (88) |

[1] Here $\eta$ shows an averaged (over a global data set) speedup or how many times NN emulation is faster than the original parameterization in a sequential single processor code by code comparison.

The NN complexity $c$ (9) and average speedup $\eta$ (how many times NN emulations are faster than the original parameterization) are also shown in Table 1. The NN complexity per output, $c$, is used because NNs with different numbers of outputs are compared here. In this case $c$ provides a more adequate metric for comparisons. For the LWR parameterization, we see a significant decrease in the speedup for NCEP CFS with 64 vertical layers vs. NCAR CAM with 26 vertical layers, although the LWR NN emulation for NCEP CFS is still 16 times faster than the original parameterization. For the SWR parameterization the opposite tendency is observed; that is, the NCEP CFS SWR NN is more than three times faster than the NCAR CAM SWR NN.

These seemingly contradictory results for the speedup achieved by LWR and SWR emulations can be explained by the interplay of the two main contributing factors: the physical and mathematical complexities of the parameterization itself (the number of treated gas species, spectral bands,, etc.), and the dependence of the particular numerical scheme implemented in the radiative transfer parameterization on the number of vertical layers in the model (for detailed discussion of this topic see Krasnopolsky et al., 2010). The results presented in Table 1 reflect the fact that the numerical scheme implemented in the NCEP CFS RRTMG-LW parameterization (Clough et al., 2005) is significantly more efficient (linear with respect to the number of vertical levels $L$) than that of the NCAR CAM LWR (Collins et al., 2002) parameterization (quadratic with respect to $L$). Thus, a smaller speedup factor is produced by the NN emulation for NCEP CFS LWR than that of NCAR CAM LWR. The NCEP CFS's RRTMG-SW includes more spectral bands and uses more complex treatment for a larger variety of absorbing/scattering species than NCAR CAM SWR; thus, NN emulation shows a larger speedup value, $\eta$, for NCEP CFS SWR than that of NCAR CAM SWR and NCEP CFS LWR. It was shown recently (Belochitski and Krasnopolsky 2021) that shallow-neural-network-based emulators of radiative transfer parameterizations developed almost a decade ago for a state-of-the-art general circulation model NCEP CFS are robust with respect to the substantial structural and parametric change in the host model: when used in two seven month-long experiments with the current significantly changed CFS, they remain stable and generate realistic output (see for example, figure 4).

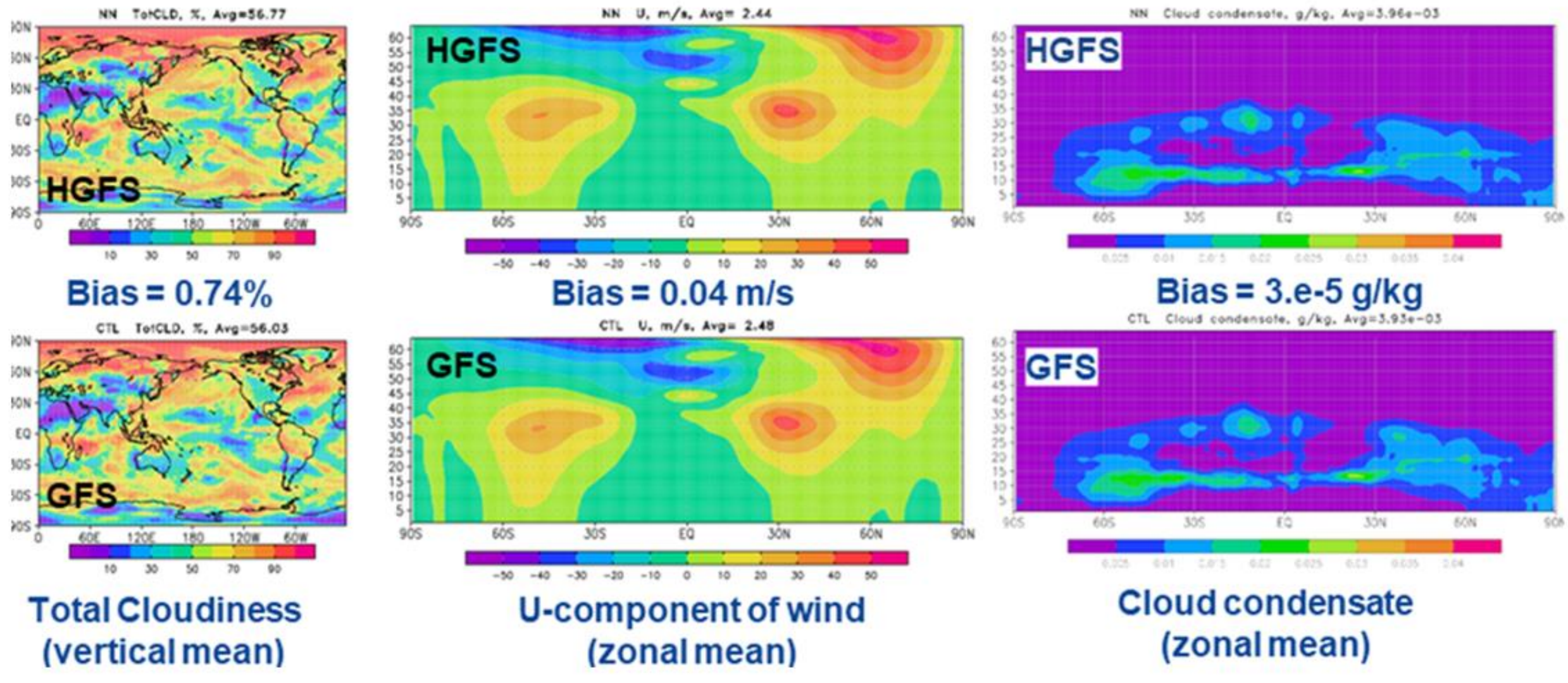

**Figure 4.** 10-day average of the total cloudiness (left), the U-component of wind (center), and the cloud condensate (right) over a forecast initialized at 0Z on 01/01/18 produced by HGFS (upper panels) and the control GFS (lower panels).

The radiative transfer calculations take a different amount of time under different cloud conditions because of the varying complexity of cloud-radiation interaction. More detailed estimations of speedup have been separately performed for three different types of cloudiness: clear sky, three cloud layers, and a more complex cloud condition where deep convection occurs. Calculation speedup for a deep convection case is shown in Table 1 in parentheses. For more complex cloud-radiation interactions in the deep convection case, the calculation of the original LWR and SWR parameterizations takes ~22% and ~57% more time, respectively, than for clear sky conditions. The time required for the NN radiation calculations is independent of the cloud conditions. As a result, the speedup achieved by the NN emulation is significantly higher for more complex cloud-radiation interactions. Additionally, using NN emulations in place of original physically-based parameterization allows for better load balancing in a parallel computational environment due to reduced idle processor time: calculations over clear sky atmospheric columns do not have to wait for cloudy columns calculations to complete.

Using the approach developed at NCEP, an emulator of RRTMG consisting of a single shallow NN that replaces both LW and SW parameterizations at once was developed at Korean Meteorological Agency for the short-range weather forecast model Korea Local Analysis and Prediction System in an idealized configuration with 39 vertical layers (Roh and Song, 2020). Inputs and outputs to RRTMG were saved on each 3 second time step of a 6 hour-long simulation of a squall line, and about 270,000 input/output pairs were randomly chosen from this data set to create training, validation, and testing sets. Dimensionality reduction was performed by removing constant inputs. Several activation functions were tested (tanh, sigmoid, softsign, arctan, linear) with hyperbolic tangent providing best overall accuracy of approximation. The emulator was two orders of magnitude faster than the original parameterization and was stable in a 6 hour-long simulation.

Two dense, fully connected, feed-forward deep-NN-based emulators with three hidden layers, one emulator per parametrization, were developed for LW and SW components of RRTMG-P for the Department of Energy's Super-Parametrized Energy Exascale Earth System Model (SP-E3SM) (Pal et al, 2019). In SP-E3SM, radiative transfer parameterizations act in individual columns of a 2-D cloud resolving model with 31 vertical levels embedded into columns of the host GCM. Calculation of cloud and aerosol optical properties were not emulated; instead, original RRTMG-P subroutines were used. Inputs and outputs of radiative parameterizations were saved on every time step of a year-long model run, with 9% of this data randomly chosen to form a data set of 12,000,000 input/output records for LW, and of 6,000,000 input/output records for SW emulator training and validation. 90% of the data in these sets was used for training, and 10% for validation and testing. No additional dimensionality reduction was performed. Sigmoid AF was chosen as it was found to provide slightly better training convergence than the hyperbolic tangent. The emulator was an order of magnitude faster than the original parametrization and was stable in a year-long run.

A number of ML-based radiative transfer parameterizations or their components have been developed, but, to our knowledge, have not yet been tested in an online setting, or in interactive coupling to an atmospheric model. Among them are deep-NN-based parameterizations of gas optical properties for RTTMG-P (Ukkonen et al, 2020; Veerman et al, 2021), and a SW radiative transfer parameterization based on convolutional deep neural networks (Lagerquist et al, 2021).

A more detailed description of the NN emulation approach can be found in Chapter 4.2 of Krasnopolsky (2013) and in Pal et al. (2019).

### 4.2 NNs emulating super parameterizations

The super-parameterization in a MMF is conceptually similar to a traditional parameterization of model physics (Krasnopolsky et al., 2014). On each time step, the embedded CRM receives a vector of input parameters $X$ from the host GCM, which describes the state of the atmosphere in a column in terms of the GCM variables. After integration of the CRM in the column of the GCM, the SP returns to the GCM a vector of output parameters $Y$, which describes the physical forcing for this column in terms/variables of the GCM. As a result, the entire SP, from a mathematical point of view, can be considered as a mapping. Given the physical and mathematical properties of the CRM, this mapping is continuous or almost continuous and can be emulated by a ML tool with a desired accuracy.

However, when it comes to emulation of SP, stochastic behavior of the parameterization has to be taken into account. This behaviour emerges due to variability associated with the internal state of the CRM, $\xi$, that are stored between the GCM time steps. The internal state is “hidden” from the host GCM environment. Because of this, given the same inputs from the GCM, the SP may generate different outputs, depending on the stored internal state, $\xi$. Hence, the SP is not completely deterministic, the SP outputs are not completely determined by the SP inputs, and the SP should be considered a stochastic mapping (2), $Y = F(X, \xi)$. As a result, uncertainty will emerge in the simulated data, which cannot be accounted for by a single emulating NN. Thus, an ensemble of NNs is better suited for emulating the SP than a single emulating NN (Krasnopolsky et al., 2014). As it could be expected, not all SP outputs demonstrate stochastic behavior to the same degree. For example, while temperature tendencies and cloud fractions are almost deterministic, water vapor and ice water tendencies clearly demonstrate a significant stochastic behavior. As a result, using an ensemble of NNs reduces approximation errors for water vapor and ice water tendencies but has almost no effect on the accuracy of temperature tendencies and cloud fractions (Krasnopolsky et al., 2014). Jones et al. (2019a,b) averaged several SP models to create a "deterministic SP". They obtained some significant differences in the climate and variability, but overall, the effect of stochasticity was limited. Rasp et al. (2018) emulated the SP using a single deep NN. Their approach does not take into account the uncertainty discussed above. They showed that using an NN speeds up calculations 20-fold. Such performance improvement opens the possibility to make experimentation with MMF substantially less computationally demanding. Also, development of SP including 3D CRM, with emulation of this SP with NN, may become feasible.

Recently, Wang et al. (2021) used a group of deep residual multilayer perceptrons with strong nonlinear fitting ability to emulate embedded CRM in an MMF configuration with realistic orography. The resulting hybrid MMF is 20 to 30 times as fast as the original reproduces reasonable climatology and climate variability and remains stable in muti-year experiments.

### 4.3 NN emulating microphysics

As mentioned above, parameterizations necessarily simplify representation of microphysical (MP) processes; as a result, applicability of a given MP parameterization may be limited to a subset of possible atmospheric scenarios. For example, Gilmore et al (2004) have found pronounced differences in the amount of predicted rainfall in numerical simulations of a deep convective storm using liquid-only vs liquid-ice MP schemes. Moreover, even when MP parameterizations of comparable complexity are used, fundamental lack of sufficient understanding of many individual processes results in differences in simulations of atmospheric phenomena, e. g. supercell thunderstorms (Morrison and Milbrandt, 2011), atmospheric rivers (Jankov et al, 2009), or tropical mesoscale convective systems (van Weverberg et al., 2013). Thus, a particular MP scheme performs well in certain atmospheric situations and performs not so well in others.

When and why one scheme outperforms others is often not well understood. It appears that none of the existing MP parameterizations may offer the comprehensive treatment of the natural processes involved. Also, even in parameterized form, MP calculations are time consuming.

Thus, ML tools can perform two different but related tasks when applied to MP parameterizations: (1) create fast ML MPs by emulating various MP parameterizations; for example, the Thompson MP scheme (Thompson, 2008) was emulated with an ensemble of NNs (Krasnopolsky et al., 2017), and (2) integrate existing MP parameterizations in a more comprehensive scheme that is able to offer better treatment of sub grid processes involved, cover greater variety of sub grid scenarios, and stochastically represent uncertainty in MP schemes.

### 4.4 NN emulating entire model physics and GCM

Building ML emulation of the entire block of atmospheric physics parameterizations in a state-of-the-art GCM is an attractive task: if successful, it could contribute to development of methodology for constructing new ML-based "empirical" parameterizations, in addition to the speedup of model calculations. This task may be aided by the fact that the full diabatic forcing profiles in atmospheric models are generally smoother than forcing profiles from individual processes, because the latter often balance each other. Krasnopolsky et al. (2009) discussed problems arising when emulating full physics using NNs and developed methods to solve some of these problems.

Several attempts have been made to create ML emulations of regional and simplified GCMs (Van der Merwe et al., 2007; Scher, 2018; Dueben and Bauer, 2018; Scher and Messori, 2019; Willard et al., 2021). It was shown by Van der Merwe et al. (2007) that NN can provide very fast (1000 times faster) and accurate emulation of a coupled large-scale circulation model for the Columbia River, its estuary, and near ocean regions. The circulation model is highly nonlinear and is driven by ocean, atmospheric, and river influences at its boundaries. The NN provides an accurate emulation of the entire model. Scher (2018) used a convolutional NN to emulate a dry hydrostatic aquaplanet at ~500 km horizontal resolution with 10 vertical levels, without either diurnal or seasonal cycles and with idealized physics. This NN used the global state of winds, temperature, and geopotential height as inputs, and predicted the global state of the same variables on the next time step. It runs stably and with reasonable realism as compared to the control for 1000 model years. In the follow up work, Scher and Messori (2019) assessed how the complexity of a climate model affects the forecast skill of the emulating NN, and how dependent the skill was on the length of the training period. They found that using neural networks to emulate the climate of a general circulation model with a seasonal cycle remained a challenge - in contrast to earlier promising results on a model without a seasonal cycle. Dueben and Bauer (2018) used a toy model for global weather prediction to identify challenges and fundamental design choices for a forecast system based on neural networks. Such fast and accurate emulations (if successful) will enable significant advances in development of geophysical modeling systems. They may serve as an improved incarnation of statistical models widely used before the numerical weather prediction era for short term and local forecasting.

# 5 Using ML to develop new parameterizations: training coarse-grid parameterization using data from fine-grid simulations

The ML techniques can also be used to improve model physics. Substantial uncertainties remain in representation of cloud properties in GCMs (e.g., Ceppi et al., 2017; Brenowitz & Bretherton, 2018; Rasp et al., 2018). Cloud Resolving Models (CRM) resolve many of the phenomena that lower resolution global and regional models do not (e.g., higher resolution fluid dynamic motions supporting updrafts and downdrafts, convective organization, meso-scale circulations, and stratiform and convective components that interact with each other, etc.).

ML techniques can be used to build a bridge or interface between CRMs and GCMs, for example, to develop a ML parameterization of moist processes, which can be used as a parameterization in GCMs and can effectively account for major sub-grid scale effects taken into account by other approaches (e.g., MMF, see 4.2). The idea is to develop an MLP, which emulates the behavior of a CRM or large eddy simulation (LES) model and can be used at larger scales (e.g., GCM scales) in a variety of regimes and initial conditions. The resulting emulation can be used as a novel, and computationally viable parameterization of sub-grid processes in a HGCM (e.g., Krasnopolsky et al., 2011, 2013; Schneider et al., 2017; Brenowitz & Bretherton, 2018; Gentine et al., 2018; O'Gorman & Dwyer, 2018), planetary boundary layer processes (Wang et al. 2019), etc. It may produce a parameterization of similar or better quality compared to the SP (that uses a simplified 2D CRM), effectively taking into account sub grid scale effects (with respect to the scales a GCM) at a fraction of the computational cost (Krasnopolsky et al., 2013, Wang, 2021). Bolton and Zanna (2019) trained convolutional NNs on data from a high-resolution quasi-geostrophic ocean model. They demonstrated that convolutional NNs successfully replicated the spatiotemporal variability of the subgrid eddy momentum forcing, could generalize to a range of dynamical behaviors, and could be forced to respect global momentum conservation.

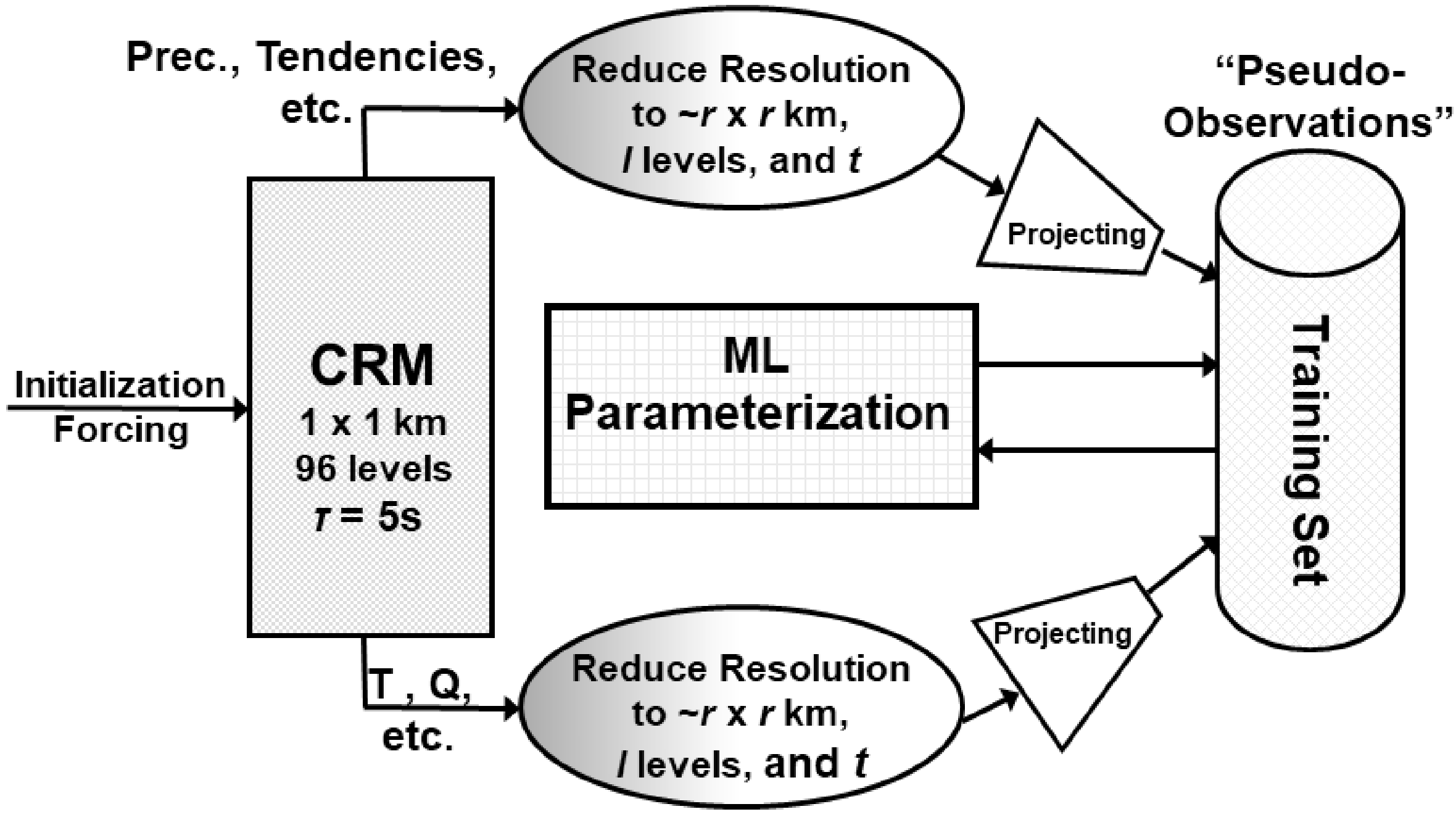

**Figure 5.** The process of preparation for training of a ML convection parameterization includes several steps: (1) simulating CRM data, (2) reducing the resolution of simulated data, (3) projecting a CRM vector of atmospheric states onto a GCM vector of atmospheric states, (4) creating training set of “pseudo-observations”

Figure 5 summarizes the process of development of an ML parameterization. The CRM uses data from, for example, the Tropical Ocean Global Atmosphere Coupled Ocean-Atmosphere Response Experiment (TOGA-COARE), Department of Energy Atmospheric Radiation Measurement Southern Great Plains observatory, or other observational campaigns, for initialization and forcing. The development of an MLP is a multi-step process. These steps are (Krasnopolsky et al., 2011):

1. *Generating CRM data*. The model is run for the period of time necessary to capture phenomena of interest, which may be limited by the data available for initialization and forcing, and the high-resolution output of the model is archived.

2. *Reducing the resolution of simulated data*. The high-resolution CRM simulated data are averaged in space and time. The data are averaged to a reduced horizontal resolution of $\rho < r \leq R$, where $\rho$ and $R$ are the CRM and GCM resolutions, correspondingly, and are interpolated/averaged to the number of vertical layers $l = L$, where $L$ is the number of vertical layers in the GCM.

3. *Projecting a vector of CRM atmospheric states onto a vector of GCM atmospheric states*. From the reduced resolution CRM data created on the previous step, the subset of variables is selected. We will refer to this subset as the NN development set. Only variables that have corresponding GCM variables or can be calculated from or converted to prognostic or diagnostic variables available in the GCM, are included in the NN development set (they are called “pseudo-observations" in Fig. 5). Only these variables can be used as inputs and outputs in the NN parameterization. In general, CRM models utilize a different set of prognostic variables than a GCM that may include, for example, turbulent kinetic energy, sophisticated treatment of precipitation species etc. If these variables are not available in the GCM, they cannot be used as MLP inputs and/or outputs. From the point of view of GCM “model reality” these variables are “hidden” variables responsible for sub-grid scale variability. The acknowledgement of this challenge requires the introduction of the concept of uncertainty and "stochasticity". The development set of pseudo-observations implicitly represents a stochastic convection parameterization (i.e., a stochastic mapping) with an uncertainty that is an inherent feature of such a parameterization.

4. *Adjusting the differences*. The pseudo-observations used for the development of the MLP are not real observations. They represent averaged CRM simulated data. Pseudo-observations are used for development of the ML parameterization that is introduced into a GCM. The GCM has its own simulated atmosphere, which may not be in complete agreement with the averaged CRM simulated data, and therefore, with the MLP trained on pseudo-observations derived from the averaged CRM simulated data. Thus, special effort may be required to synchronize or make consistent the atmospheric realities of the GCM and the averaged CRM simulated data. CRM and the GCM mean differences for all variables selected as MLP inputs and outputs must be determined and compensated for (Krasnopolsky et al., 2011). These differences are the result of GCM and CRM being two different models with different temporal and spatial scales and resolutions, with different dynamics and physics; they also have different boundary and initial conditions and different forcing.

5. *Creating data sets*. The development set of pseudo-observations is separated into the independent training and test/validation data sets. Then the MLP is trained using the training set. Due to the inherent uncertainty of pseudo-observations, the MLP derived from these data is a stochastic parameterization. One of the ways to take the stochasticity into account is to implement the MLP as an ensemble of NNs (Krasnopolsky et al., 2011, 2013; see also the next Section).

6. The validation procedure for the MLP consists of two steps. First, the trained ensemble stochastic MLP is applied to the test set and error statistics are calculated. Second, the tested MLP is introduced into the GCM to validate its behavior in the model simulations. *This last step is the most important step of the validation process.*

The same procedure can be applied for developing parameterization of subgrid processes for any coarse resolution model using a fine resolution model. For example, an LES model can be used to generate data for the following application of a ML tool. Another approach that was introduced recently proposes to use the existing DA system to learn aspects of the model by doing parameter estimation, or by learning new functional representations using neural networks or another ML approach (Bocquet et al., 2019). Such an approach of learning within an existing data assimilation system is feasible for a weather forecasting center that is already operating a data assimilation system that encodes a wealth of weather forecasting knowledge accumulated during decades (Geer 2021).

ML can also be used for improving existing parameterization without following hybridization of the model. For example, genetic algorithms were used to optimize parameters of generalized multiple discrete interaction approximation for wind waves models (Tolman and Grumbine, 2013; Tolman, 2013). The free parameters of the parameterization are optimized holistically, by optimizing full model behavior in the WAVEWATCH wave model (Tolman, 2009). Then this optimized parameterization was used in the WAVEWATCH model and resulted in improved forecasts especially in extreme weather conditions.

# 6 Using ensemble of ML tools

An ensemble of statistical models is usually used when a single model cannot unambiguously account for the variability of the data. Such a situation may occur when, for example: (i) the data contain a significant amount of noise, and deriving an ensemble of models with the following averaging of the ensemble reduce the influence of the noise; (2) the data contain an inherent stochastic component and can be considered as representing a family of functions, as in the case of a stochastic function or mapping; (3) the problem is ill posed and a single statistical model does not give stable results, as in the case of nonlinear extrapolation or calculating derivatives of statistical model.

## 6.1 Stochastic parameterizations and stochastization of deterministic parameterizations

As was mentioned above, the stochastic parameterization mapping (2) may be considered a family of mappings distributed with a probability density function. Each mapping of the family may be considered to be one possible realization of subgrid processes. The range and shape of the distribution function are determined by the uncertainty vector ε. An average over the subgrid scenarios represented by the stochastic parameterization can be approximated by a single ML tool, e. g. a single NN. However, an ensemble of ML tools can give a better and more adequate representation of the stochastic parameterization, provided each member of the ML ensemble emulates one or a few functions of the family (one or a few subgrid scenarios).

The ensemble of ML emulations can also be used for stochastization of deterministic parameterizations to create perturbed physics ensembles (Krasnopolsky et al., 2008b). Usually perturbed physics (or parameterization) is created by adding a small random value to deterministic physics. When ensembles of ML emulations are used, a particular emulation (an ensemble member number j) may be considered to be the $j^{th}$ perturbed version of the unperturbed model physics (or parameterization), $P$, and can be written as,

$$P_j^{ML} = P + \varepsilon_j \tag{13}$$

where $P_j^{ML}$ is a ML emulation (ensemble member) number $j$ of the original model physics, $P$, and $\varepsilon_j$ is the approximation error for the ML emulation number $j$. If an SNN is used for ML emulation, both the mean value and statistical properties of $\varepsilon_j$ can be controlled and altered significantly by varying the number of hidden neurons $k$ in eq. (6) (Krasnopolsky & Fox-Rabinovitz, 2006a). For example, the systematic component of the approximation error (bias) can be made negligible and, therefore, $\varepsilon_j$ almost purely random. Thus, $\varepsilon_j$ can be made the same order of magnitude as the natural uncertainty of the model physics (or of a particular parameterization) due to the unaccounted variability of sub-grid processes. A single ML emulation can be considered a stochastic version of the original deterministic scheme.

## 6.2 Calculating Jacobian of a parameterization

In some applications, first derivatives of MLP are required, e.g., for calculation of an adjoint of an ML parameterization in a 4-D variational data assimilation system (Chantry et al., 2020) or for error and sensitivity analysis. It means that the ML emulation Jacobian, *J*, that is a matrix of the first derivatives of the outputs of the ML emulation (e.g., 6) over its inputs has to be calculated,

$$J = \left[\frac{\partial y_j}{\partial x_p}\right]_{p\,=\,1,\dots,N}^{j=1\,,\dots,M} \tag{14}$$

More specifically, for SNN,

$$\frac{\partial y_j}{\partial x_p} = \sum_{i=1}^{k} a_{ji}^{1} \cdot \left(1 - t_i^2\right) \cdot a_{ip}^{0}$$

After substituting (6) into the above expression, a simple equation for elements of the Jacobian matrix (14) for an SNN can be obtained,

$$\frac{\partial y_j}{\partial x_p} = \sum_{i=1}^{k} a_{ji}^{1} \cdot \left\{1 - \left[tanh\left(b_i^0 + \sum_{s=1}^{N} a_{is}^{0} \cdot x_s\right)\right]^2\right\} \cdot a_{ip}^{0}$$

However, from a theoretical point of view, inference of the NN Jacobian is an ill posed problem (Vapnik, 1995), which in practice leads to significant uncertainties in NN Jacobians calculated in a manner described above (Aires et al., 1999; Aires et al., 2004; Chevallier & Mahfouf, 2001). Several solutions have been proposed to reduce these uncertainties. First, if a data set for a Jacobian is available, the Jacobian can be trained as a separate additional NN (Krasnopolsky et al., 2002). Alternatively, the Jacobian can be included as additional outputs of the NN emulation. To do this, the error (or cost) function, which is minimized in the process of NN training, should be modified accordingly: the Euclidean norm, which is usually used in calculating the error function, should be replaced with the first order Sobolev norm. Actually, it was shown that the function of the Sobolev space can be approximated by a NN with all their derivatives (Hornik et al., 1990).

Second, the mean Jacobian (over time or space) can be calculated and used (Chevallier & Mahfouf, 2001). Third, regularization techniques like "weight smoothing" (Aires et al., 1999) or a technique based on a principal component decomposition (Aires et al., 2004) can be used to stabilize the Jacobian. Finally, an MLP Jacobian can be calculated using the NN ensemble approach (Krasnopolsky, 2007a), which allows one to calculate the ML emulation Jacobian with an accuracy sufficient for practical applications. The ensemble approach: (a) significantly reduces the systematic and random error in ML emulation Jacobian, (b) significantly reduces the magnitudes of the extreme outliers and, (c) in general, significantly reduces the number of larger errors. In the NN ensemble approach, an ensemble of $q$ ML emulations of a parameterization is trained. Then, for each specific ML emulation a partial Jacobian, $J_i$, is calculated and the Jacobian, $J$, in turn, is calculated as an average of partial Jacobians:

$$J = \frac{1}{q}\sum_{i=1}^{q} J_i$$

# 7 Ensuring physical constraints

Physical constraints/relationships, such as energy conservation, are embedded in the model. The physically based parameterizations preserve physical invariants (energy, momentum, etc.) with high accuracy because these relationships are explicitly (or implicitly) built into the parameterizations. MLPs usually preserve physical invariants only approximately but can approximate physical constraints with high accuracy (Krasnopolsky et al. 2008c; O'Gorman & Dwyer 2018). For example, for the random forest (RF) convection parameterization developed by O'Gorman and Dwyer (2018), the root-mean-squared error (RMSE) in conservation of column-integrated moist enthalpy in the control climate is very small at 0.2 W $m^{-2}$ for both the training dataset and the RF predictions on the test dataset. For the NN emulations of radiation parameterizations developed by Krasnopolsky et al. (2010), the mean error does not exceed $6.5 \cdot 10^{-4}$ K $day^{-1}$. Both these errors do not cause any significant deviation from the control run in parallel run validations. In NCEP CFS parallel runs with NN LWR and NN SWR, which were run for 17 years, the deviation from the balance was corrected using the following approach (Krasnopolsky et al., 2010). The integral relationship for the imbalance, $\varepsilon$, that relates pressure, heating rates, and fluxes was used,

$$\varepsilon = \frac{\sum_{k=1}^{L} \alpha_k \cdot h_k}{\sum_{k=1}^{L} \alpha_k} + \frac{\Phi}{\sum_{k=1}^{L} \alpha_k} = 0.$$

$$\Phi = \begin{cases} F_{tup} - F_{\sup} + F_{sdn} & for\ LWR \\ F_{tup} - F_{tdn} - F_{\sup} + F_{sdn} & for\ SWR \end{cases} \tag{15}$$

where, $\alpha_k = \frac{p_k - p_{k-1}}{G}$, $p_k$ and $h_k$ are pressure and heating rates at a vertical level $k$, $G$ is a constant, $F_{tup}$ is the total sky outgoing LWR or SWR flux at the top of the atmosphere, $F_{tdn}$ is the total sky downward SWR flux at the top of the atmosphere, $F_{sup}$ is the total sky upward LWR or SWR flux at the surface, and $F_{sdn}$ is the total sky downward LWR or SWR flux at the surface.

The outputs of the original radiation parameterizations satisfy the relationship (15) with high accuracy because these relationships are explicitly (or implicitly) included into the parameterizations. The outputs of the NN emulations satisfy (15) only approximately, i.e., in this case, the imbalance $\varepsilon \neq 0$; $\varepsilon$ however, is small. For example, for the RRTMG – NN LWR emulation, mean value for $\varepsilon$ is $6.5 \cdot 10^{-4}$ K/Day.

A correction can be introduced for the heating rates, $h_k$. The correction makes the modified or balanced heating rates, $\widetilde{h_k} = h_k + \varepsilon$, to satisfy the relationship (15) exactly. This correction is very small and, as a result, the balancing procedure does not practically affect the overall accuracy of LWR NN, marginally improves the overall accuracy of SWR NN, and does not change the results of parallel runs (Krasnopolsky et al., 2010).

In general, ML emulations that are trained using data simulated by a model and simply mimic the existing parameterization may not require additional physical constraints as these constraints are implicitly incorporated in the simulated data that have very low levels of noise. If the training set is representative enough, which is comparatively easy to achieve with simulated data, ML emulations can learn physical constraints from the data with sufficient accuracy, as illustrated by the results of Krasnopolsky et al. (2010)

outlined above. On the other hand, when a new ML parameterization is developed using observed data or data simulated by finer resolution models, the fact that many sources of noise and uncertainties contribute to the training data has to be taken into account. In this case applying physical constraints during the training may be the only way to avoid instabilities after the trained ML parameterization is incorporated into a hybrid model.

Several approaches can be used to enforce physical constraints. ML techniques that explicitly use training set values to calculate predicted values for the new inputs, e. g., trees and SVMs, will preserve physical constraints with the same degree of accuracy as the data comprising the training set (O'Gorman & Dwyer, 2018). However, these techniques are known to be less accurate when generalizing to data do not present in the training set. Beucler et al. (2019) proposed two methods to enforce energy conservation laws in neural-network emulators of physical models, (1) modifying the loss function to include energy conservation laws, or (2) constraining the architecture of the network itself. They showed that when applied to the emulation of explicitly resolved cloud processes in an idealized multi-scale climate model, the architecture constraints can enforce conservation laws to satisfactory numerical precision, while adding constraints on the loss function helps the NN to better generalize to conditions outside of its training set.

# 8 Going beyond the traditional paradigm

## 8.1 Compound parameterization to improve robustness of MLPs

The accuracy of MLPs depends significantly on our ability to generate a representative training set and to avoid as much as possible using MLPs for extrapolation beyond the domain covered by its training data. Because of high dimensionality of the input domain (i.e., dimensionality of the MLP input vector X), which is usually of the order of several hundred or more, it is difficult if not impossible to cover the entire domain, which may have a very complex shape, even when we use model simulated data for the MLP training. Also, the domain may change with the structural and parametric modifications to the host model, e.g., change or replacement of the dynamical core or other model components, change in horizontal resolution, or chemical composition of the atmosphere due to climate change. In these situations, the MLP may be forced to extrapolate beyond its generalization ability, which may lead to larger errors in MLP outputs and, correspondingly, in numerical model simulations employing the MLP.

NN emulations of model radiation (Krasnopolsky et al., 2010) are very accurate. Larger errors and outliers (extreme errors) in NN emulation outputs occur only when NN emulations are exposed to inputs not represented sufficiently in the training set. These errors have a very low probability and are distributed randomly in space and time. However, when long multi-decadal climate simulations are performed, and NN emulations are used in a very complex and nonlinear climate model over long integration times, the probability of larger errors and their undesirable impact on the model simulations increases. As demonstrated in Krasnopolsky et al. (2008a, 2010), a GCM is robust enough to overcome rare large randomly distributed errors without them accumulating over time. In another application of a NN approximation for nonlinear interactions in a wave model, the model did not prove sufficiently robust to retain stability for a sufficient integration time (Krasnopolsky et al., 2008c). Therefore, for some applications of MLPs, it is essential to introduce a quality control (QC) procedure, which can predict and eliminate larger errors from MLP during the integration of highly nonlinear numerical models.

Krasnopolsky et al. (2008c) introduced the concept of compound parameterization (CP) that combines an MLP, the original physically-based parameterization, and a QC procedure. CP makes the MLP emulation approach even more reliable, robust, and generic. It also provides a tool for developing MLP emulations adjustable to changes in the structural and parametric changes in the host model. An effective QC design is based on training an additional NN to specifically predict the errors of the MLP outputs for a given input. The "error" NN has the same inputs as the MLP and one or several outputs, which predict errors of outputs generated by the MLP for these inputs. The original parameterization, its NN emulation (MLP), the error NN, and the QC block constitute the CP, the design of which is shown in Figure 6.

During the model integration, CP works in the following way: if the error predicted by the error predicting ML tool (error NN) does not exceed a predefined threshold value, $\varepsilon_{th}$, the MLP is used; otherwise, the original parameterization is called instead of the MLP. The auxiliary training set (ATS) is updated each time when QC invokes the original parameterization instead of the MLP. ATS is used for the following adjustments of the MLP. CP can be used for an online adjustment of MLP. After each update of the ATS, the new record can be used to update the coefficients of MLP, using a sequential training algorithm.

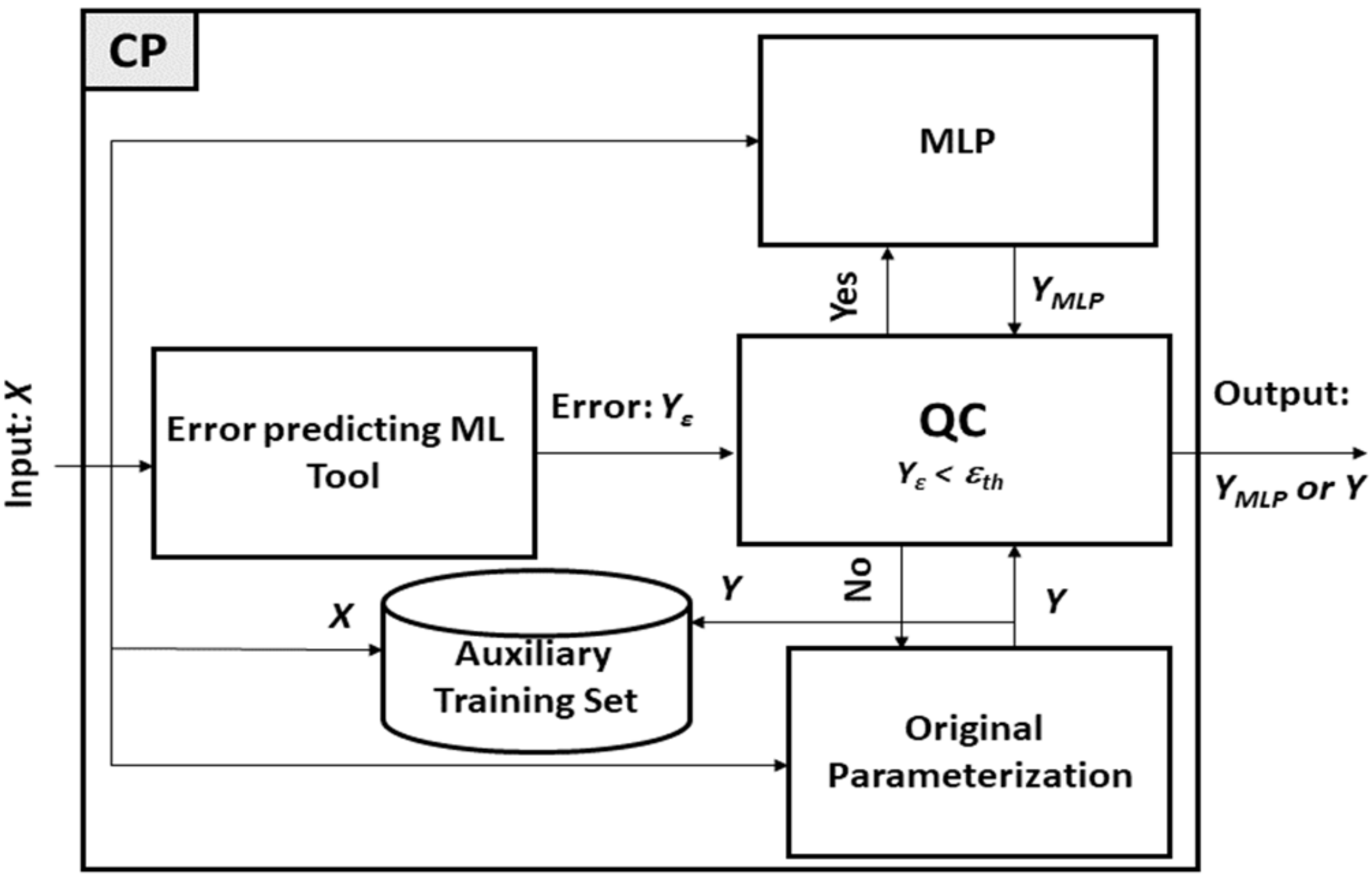

**Figure 6.** A compound parameterization design.

As it was shown in Krasnopolsky et al. (2008c), errors predicted by the error NN are close to and highly correlated with the actual errors of the MLP calculated for the same input vector. The use of a CP: (a) does not increase the systematic MLP error (bias) which is almost zero; and (b) significantly reduces random error of the MLP. Especially significant is the reduction of extreme errors or outliers. When the CP approach was applied to ML emulation of the SWR parameterization, only fewer than 1% of the MLP outputs were rejected by the QC and re-calculated using the original parameterization (Krasnopolsky et al., 2008c). Other designs of CP can be developed and used. For example, a design based on the combination of forward and inverse NNs was successfully applied for the nonlinear wave-wave interaction parameterization in a wind wave model (Krasnopolsky et al., 2008c). Recently Song et al. (2021) applied the CP design proposed by Krasnopolsky et al. (2008c) to the ML emulations of both LW and SW radiative transfer parameterizations. They compared it to a CP scheme that uses vertically integrated cloud fraction instead of the full set of MLP inputs in its QC component and found the latter to be more accurate and computationally efficient.

### 8.2 Adjusting MLP to changes in the environment

After the MLP is trained and incorporated into the host model, it can successfully function in the online mode, replacing the original parameterization, as long as the input domain (domain spanned by MLP inputs provided by the model) does not change significantly. The domain may change with the evolution of the environment, both internal model environment (change of dynamical core, model horizontal resolution, other parameterizations, etc.) and external environment (the system that this model is simulating, e.g., due to climate changes). With domain changes the accuracy of MLP can deteriorate when it encounters input vectors, $X$, underrepresented or not represented at all in the training set. To mitigate this problem, a concept of dynamical adjustment of MLP (Krasnopolsky & Fox-Rabinovitz, 2006a,b) was introduced. When MLP outputs are rejected by the QC block of the CP described in the previous subsection (Fig. 6), the original parameterization inputs and outputs are accumulated in the auxiliary training set, augmenting the data underrepresented in the originally generated training set. The auxiliary training set can be used to further adjust the NN emulation. After accumulating a sufficient number of records, an adjustment of the MLP emulation is performed by a short MLP retraining, using the accumulated input/output records.

The MLP can be updated directly in the model online with addition of each new record to the auxiliary training set using a one iteration of sequential training algorithm (see Krasnopolsky, 2013, Chapter 2). As a result, the MLP is dynamically adjusted online to the changes and/or new events/states produced by the host model. After a period of dynamical adjustments, the MLP will benefit from full retraining with new data. Dynamically adjusted MLP can be used during the transitional period from old MLP to a new completely re-trained MLP, allowing enough time to collect a new training set. Rasp (2020) developed a similar concept of "coupled online learning", using the high-resolution simulation that is kept in sync with the neural network-driven lower-resolution model through continuous nudging.

# 9 Summary and Discussion

In this overview, we discussed using ML approaches in weather and climate numerical modeling systems. We mentioned here two generic types of ML applications to model physics:

1. ML emulations of existing parameterizations for speeding up the model integration process

2. Development of new MLP to improve the accuracy of representation of physical processes in environmental models

A new type of model based on a synergetic combination of deterministic modeling with ML within a hybrid model is introduced. This approach uses ML tools (like NNs) to develop new MLPs and/or highly accurate and fast emulations of existing model physics. Advantages and limitations of the ML technique (and the NN technique in particular) are discussed in detail in Chapters 2 and 4 of Krasnopolsky (2013). Here we will mention only those that are relevant for development of MLPs.

## 9.1 Possible advantages of using MLPs

The results presented in this overview and cited papers show that:

(i) There exists the conceptual and practical possibility of developing hybrid models with accurate and fast MLPs as model components, which preserve the integrity and all the detailed features in the original model (e.g., GCM). ML components can be successfully combined with deterministic model components within the hybrid model, so their synergy can be efficiently used for environmental and climate modeling without any negative impacts on simulation quality.

(ii) ML emulations of existing model physics parameterizations are functionally identical to the original physical parameterization, due to the capability of ML techniques to accurately approximate complex mappings like parameterizations of the model physics. This capability allows the integrity and level of functional complexity of the parameterizations of the model physics to be preserved. As a result, a hybrid model, using these MLPs, produces climate simulations and weather forecasts that are practically identical to those of the original GCM. It is noteworthy that the MLPs have the same inputs and outputs as the original parameterizations and are used precisely as their functional substitutes within the model.

(iii) Accurate MLPs are robust and very fast (10 to $10^5$ times faster than the original parameterization) so the significant speed-up of the model calculations can be achieved without compromising accuracy.

(iv) New computationally efficient MLPs can be developed based on learning from observed data, analysis, and/or data simulated by higher resolution models.

(v) The productive synergy of state-of- the-art deterministic and ML approaches leads to new opportunities for using hybrid models for environmental and climate simulations and weather prediction. For example, new more sophisticated parameterizations, or even “superparameterizations” using CRM or LES, that are extremely time consuming or even computationally prohibitive if used in their original form, will become computationally “affordable” in hybrid models when using their accurate and much more computationally efficient ML emulations.

(vi) The stochastic nature of some of the components in the model physics can be adequately represented using ensembles of ML tools to represent or emulate the stochastic component of the model physics.

(vii) ML techniques are very flexible; they could provide an opportunity to include in consideration important variables that have not been included in physically based parameterization. In MLPs these variables can be included as additional inputs.

(viii) In principle, MLPs can go beyond the single column physics paradigm and provide a useful tool to include in model physics interaction between neighboring vertical columns, producing ML 3-D model physics.

## 9.2 Stability of MLPs, limitations of the current hybrid modeling framework, and possible solutions

Development of MLPs depends significantly on our ability to generate a representative training set to avoid using NNs for extrapolation beyond the domain covered by the training data. Because of the high dimensionality of the input domain that is often several hundred or more, it is rather difficult to cover the entire domain, especially the "far corners" associated with rare events, even when using simulated data for MLP training.

Representativeness of a training set is so important because, as is the case with any nonlinear statistical tool, ML techniques are not very good at generalization, especially extrapolation; nonlinear extrapolation is an ill-posed problem that requires a regularization to provide meaningful results. Poor generalization may lead to instability of the emulation in the host hybrid model. The ML ensemble approach could be of significant help here. Using an ensemble of ML tools can help to regularize the extrapolation and result in MLPs that are more stable when the inputs approach or cross the boundary of the training domain.

Developing a stable and robust NN-based emulator is a multifaceted problem that requires deep understanding of multiple technical aspects of the training process and details of NN architecture. Many techniques for stabilization of hybrid statistical-deterministic models have been developed. Compound parameterization has been proposed for climate and weather modeling applications where an additional NN is trained to predict errors of the NN emulator, and, if the predicted error is above a certain threshold, compound parameterization falls back to calling the original physically-based scheme (Krasnopolsky et al, 2008). Stability theory was used to identify the causes and conditions for instabilities in ML parameterizations of moist convection when coupled to an idealized linear model of atmospheric dynamics (Brenowitz et al, 2020). An NN optimization via random search over hyperparameter space resulted in considerable improvements in stability of subgrid physics emulators in the Super-parameterized Community Atmospheric Model version 3.0 (Ott et al, 2020). A coupled online learning approach was proposed where a high-resolution simulation is nudged to the output of a parallel lower-resolution hybrid model run, and the ML-component of the latter is retrained to emulate tendencies of the former, helping to eliminate biases and unstable feedback loops (Rasp, 2020). Random forests approach was successfully used to build a stable ML parameterization of convection (Yuval and O'Gorman, 2020). Physical constraints were used to achieve stability of hybrid models (e.g., Yuval et al, 2021; Kashinath et al, 2021).

Reusing methodologies from one ML field in another is often suggested as a particularly powerful tool (e.g., Boukabara et al, 2019), allowing to leverage existing knowledge developed in other ML applications. One example of such an approach is transfer learning. However, this powerful tool should be used carefully. ML has been applied to many different problems such as image processing, classification, pattern recognition, asynchronous signal processing, feature detection, data fusion, merging, morphing, mapping, etc. From the mathematical standpoint, these problems belong to different mathematical classes. Applying a methodology developed for a problem belonging to a mathematical class different from the class of the problem at hand may not necessarily be justified. Therefore, caution should be exercised when applying techniques designed for problems other than mapping approximation to development of model physics emulators or new ML-based parameterizations. One of such very popular techniques is deep learning.

Application of shallow NNs to the problem of mapping approximation has thorough theoretical support. The universal approximation theorem proves that an SNN is a generic and universal tool for approximating any continuous and almost continuous mappings under very broad assumptions and for a wide class of activation functions (e.g., Hornik et al, 1990; Hornik, 1991). Similarly broad results for deep NNs (DNNs) do not exist as of yet (Vapnik, 2019), however specific combinations of DNN architectures and activation functions have theoretical support (e.g., Leshno et al, 1993; Lu et al, 2017; Elbrachter et al 2020). Until there is a universal theory, it has been suggested to consider DNN a heuristic approach since, in general, "from the theoretical point of view, deep networks cannot guarantee a solution of any selection problem that constitutes a complete learning problem" (Vapnik, 2019). These considerations are important to keep in mind when selecting NN architecture for the emulation of model physics or its components

Next, we compare some properties of DNNs and SNNs to further emphasize their differences and to point out some properties of DNNs that may lead to instabilities in deterministic models coupled to DNN-based model components.

To avoid overfitting and instability, complexity, and nonlinearity of approximating/emulating NN should not exceed complexity and nonlinearity of the mapping to be approximated. The complexity of the SNN (8) increases linearly with the number of neurons in the hidden layer, k. For given numbers of inputs and outputs there is only one SNN architecture/configuration with a specified complexity $C$. The complexity of the DNN (7) increases geometrically with the increasing number of layers, n.

Both (8) and (7) are simply numbers of parameters of the NN that are trained or fit during SNN/DNN training. While there exists a one-to-one correspondence between the SNN complexity and the SNN architecture, given the fixed number of neurons in the input and output layers, correspondence between the DNN complexity and the DNN architecture is multivalued: many different DNN architectures/configurations have the same complexity $C$ given the same size of input and output layers. Overall, controlling the complexity of DNNs is more difficult than controlling the complexity of SNN.

For an SNN given by the expression (6) nonlinearity increases arithmetically or linearly with addition of each new hidden neuron, $t_i = \phi\left(b_i^0 + \sum_{s=1}^{n} a_{is}^0 \cdot x_s\right)$, to the single hidden layer of the NN.

For a DNN, symbolically written as

$$Y = X^{n+1} = B^n + A^n \cdot \phi\left(B^{n-1} + A^{n-1} \cdot \phi\left(B^{n-2} + A^{n-2} \cdot \phi\left(B^{n-3} + \cdots \phi\left(B^0 + A^0 \cdot X\right)\right)\right)\right),$$

each new hidden layer/neuron introduces additional nonlinearity on top of nonlinearities of the previous hidden layers; thus, the nonlinearity of the DNN increases geometrically with addition of new hidden layers, much quicker than the nonlinearity of the SNN. Thus, controlling nonlinearity of DNNs is more difficult than controlling nonlinearity of SNNs. The higher the nonlinearity of the model the more unstable and unpredictable generalization is (especially nonlinear extrapolation that is an ill-posed problem).

DNNs are a very powerful and flexible technique that is extensively used for emulation of model physics and its components (Kasim et al, 2020). Discussion of its limitations can be found in Thompson et al. (2020). The arguments listed here are intended to point out possible sources of instability of DNNs in the models and the need for careful handling of this very sensitive tool.

Activation Function. Universal approximation theorem for SNNs is satisfied for a wide class of bounded, non-linear AFs. Note that many popular AFs used in DNN applications, e. g. variants of ReLU, do not belong to this class. However, for a specific problem of mapping approximation, it may be useful to consider additional restrictions on AFs.

If the AF is almost continuous, or, in other words, has only finite discontinuities (e. g., step function), the first derivative (Jacobian) of the NN using this AF will be singular. If the AF is not continuously differentiable (e. g., ReLU), its first derivative will not be continuous (will have finite discontinuities), and so will be the NN Jacobian. Using a non-continuously differentiable NN as a model component may lead to instability, especially if the Jacobian of this component is calculated in the model. Using gradient-based optimization algorithms for training such NNs may be challenging due to discontinuities in gradients.

If the AF is monotonic, the error surface associated with a single-layer model is guaranteed to be convex, simplifying the training process (Wu, 2009). When AF approximates identity function near the origin (i.e., $\phi(0) = 0, \phi'(0) = 1$, and $\phi'$ is continuous at 0), the neural network will learn efficiently when its weights are initialized with small random values. When the activation function does not approximate identity near the origin, special care must be used when initializing the weights (Susillo and Abbott, 2014).

It is noteworthy that the sigmoid and hyperbolic tangent AFs, popular in SNN applications, meet all aforementioned criteria. Additionally, in the context of emulation of model physics parameterizations, these AFs provide one of the lowest training losses, as compared to other AFs (Chantry et al, 2020).

Another problem arises when ML emulations are developed for a non-stationary environment or weather/climate system that changes with time. This means that the domain configuration for a climate simulation may evolve when using, for example, a future climate change scenario. Thus, the MLP may be forced to extrapolate beyond its generalization ability, leading to errors in MLP outputs and resulting in simulation errors and instability of the corresponding hybrid model. In this situation CP and dynamical adjustment as well as using the ML ensemble approach could be helpful.

It is noteworthy that ML still requires human expertise to succeed. As was mentioned above, the development of MLPs is not a standard ML problem. While MLPs can, in principle, be used as a black box, the development of ML physics will require domain knowledge about the Earth system. Close collaborations between computer scientists, atmospheric physicists, and modelers will be essential even if petabytes of training data and GPU supercomputers are available. A deep understanding of how to use physical knowledge of the Earth system and the connectivity between degrees of freedom to improve the development of network architectures and network training and how to preserve conservation properties will be required. There are a lot of decisions that must be made in the process of developing MLPs that cannot be made automatically. Selecting and normalizing inputs and outputs, preparing training sets, selecting a training algorithm and its parameters, making decisions about sufficient approximation accuracy, etc. requires active human participation. This creates some problems with using a commercial ML software for MLP development. For example, for MLP development a very specific normalization of outputs or rearrangement of the loss function (e.g., in the case of missed outputs or physical constraints) are sometimes required, which are not always possible with a commercial ML software.

### 9.3 Concluding remarks

NNs are the ML tool that has been used in the majority of published literature on MLPs. As mentioned in Section 3.1, SNNs are backed by a well-established mathematical theory and have a long history of successful applications. DNNs are a very powerful and flexible technique that has demonstrated its power in many applications (Kasim, 2020), including developments of MLPs (e.g., Rasp et al. 2018). However, when it comes to approximation of mappings, deep learning should currently be considered a heuristic approach (Vapnik 2019), as only specific combinations of DNN architectures and activation functions have been demonstrated to be solutions to the approximation problem. Discussion of limitations of the DNN technique can be found in Thompson et al. (2020). More than 1,000 papers and other sources were analyzed by Thompson and coauthors to understand how deep learning performance depends on computational power in different application domains. They showed that computational requirements have escalated rapidly in each of these domains and that these increases in computing power requirements have been central to performance improvements. They conclude that if progress continues along current lines, these computational requirements will rapidly become technically and economically prohibitive. Thus, their analysis suggests that deep learning progress will be constrained by its computational requirements and that the machine learning community will be pushed to either dramatically increase the efficiency of deep learning or to move to more computationally efficient machine learning techniques.

Deep learning is computationally expensive not by accident, but by design. The same flexibility that makes it excellent at modeling diverse phenomena and outperforming expert models also makes it dramatically more complex and computationally expensive. Equation (7) shows how quickly the complexity of DNN (the number of DNN weights and the dimensionality of DNN training space) increases with the increase in the number of hidden layers. This rapidly increasing number of weights has to be matched by the corresponding increase in size of the training set. Additionally, increased complexity of DNN leads to increase in nonlinearity of the DNN model. With each additional hidden layer, DNN becomes more and more nonlinear, and prediction and generalization (extrapolation as well as interpolation) using this DNN may become erroneous and unstable (e.g., Rasp 2019). Understanding of these limitations is necessary when DNN is used for developing MLPs. Functional complexity of the ML tool should correspond to the functional complexity of the modeled physical process. The parsimony principle is valid in application to ML.

In a few papers mentioned in this review, various kinds of tree algorithms were explored for MLP development. Belochitski et al (2011) compared an SNN with decision tree algorithms (nearest neighbors, regression trees, and sparse occupancy trees) in application to emulation of an existing radiation parameterization and have shown that the NN algorithm performs better than tree-based algorithms in terms of accuracy and speedup. In another study, O'Gorman and Dwyer (2018) used the random forest (RF) algorithm that consists of an ensemble of decision trees. Each tree makes predictions, using a subset of the training data. The final prediction from the RF is the average over all trees. RF has the advantage that it automatically ensures conservation of energy and positive definiteness of surface precipitation because it always produces the output that is an average of the training data that it uses. On the other hand, this property of the tree (including RT) algorithms leads to the inability of tree algorithms to generalize, to produce an output outside the domain covered by the training data, whereas NN is capable of limited generalization. However, the same property makes tree algorithms more stable than NNs when integrated in the model because to any unexpected input they respond with the output inside the training domain. O'Gorman and Dwyer (2018) showed that when attempting to derive a new parameterization of subgrid processes based on a SNN, NN did not conserve energy and was slightly less accurate than the tree algorithm. An advantage of NN as compared to tree algorithms is that the training set is used only during the training process, which is a time and memory consuming operation. After a NN is trained, its application is fast and does not require the use of training or test data sets. The only data that are required are NN weights. With tree algorithms, each application of the trained algorithm involves using the training set, which is a time and memory consuming operation, significantly reducing the speedup that this MLP provides. A comparatively small number of publications where ML tools other than NN are used in geoscientific modeling is not sufficient to make well-supported comparisons of NNs with other ML tools and to draw any conclusions about their respective comparative benefits.

It is noteworthy that, in addition to developing MLPs, ML tools can be successfully used in many other parts of the numerical weather prediction and climate simulation systems. In data assimilation systems, fast ML emulations of forward models can be used for direct assimilation of satellite measurements (Krasnopolsky, 2007b). ML observation operators were used to propagate surface observations vertically (Krasnopolsky, 2013). Also, MLPs are well suited for calculating adjoints and tangent-linear models for 4D-Var data assimilation (Chantry et al., 2020). In satellite remote sensing, ML algorithms that simultaneously retrieve several geophysical parameters from satellite data, fast ML forward models for using in retrieval algorithms and data assimilation systems (Krasnopolsky, 2008d), and a ML algorithm for filling gaps in satellite data (Krasnopolsky et al., 2016) have been developed. For post processing of the model output (Haupt et al. 2021), ML tools were used to average a multi-model ensemble for precipitation (Krasnopolsky & Lin, 2012b) and for averaging wind wave model ensembles (Campos et al., 2018). Also, an ML based biological model for ocean color has been developed to enable the feedback between physical and biological processes in the upper ocean (Krasnopolsky et al., 2018).